\documentclass[graybox,a4paper]{svmult}

\usepackage{mathptmx}       
\usepackage{helvet}         
\usepackage{courier}        
\usepackage{type1cm}        
\usepackage{makeidx}         
\usepackage{graphicx}        
\usepackage{multicol}        
\usepackage[bottom]{footmisc}
\usepackage[caption=false]{subfig}
\usepackage{todonotes}

\usepackage{amsmath}
\usepackage[mathscr]{euscript}
\usepackage{bm}
\usepackage{bbold}
\usepackage{url}
\usepackage[naturalnames]{hyperref}

\makeindex             

\newcommand{\q}[1]{``#1''}

\DeclareMathOperator*{\argmax}{argmax}

\begin{document}

\title*{Building nonparametric $n$-body force fields using Gaussian process regression}



\author{Aldo Glielmo, Claudio Zeni, \'Ad\'am Fekete and Alessandro De Vita}

\institute{Aldo Glielmo \at Department of Physics, King's College London, Strand, London WC2R 2LS, \email{aldo.glielmo@kcl.ac.uk}
\and Claudio Zeni \at Department of Physics, King's College London, Strand, London WC2R 2LS, \email{claudio.zeni@kcl.ac.uk}
\and \'Ad\'am Fekete \at Department of Physics, King's College London, Strand, London WC2R 2LS
\and Alessandro De Vita \at Department of Physics, King's College London, Strand, London WC2R 2LS; Dipartimento di Ingegneria e Architettura, University of Trieste, Piazzale Europa, 1, 34127 Trieste
}
%
%
\maketitle

\abstract{
Constructing a classical potential suited to simulate a given atomic system is a remarkably difficult task. 
This chapter presents a framework under which this problem can be tackled, based on the Bayesian construction of nonparametric force fields of a given order using Gaussian process (GP) priors. 
The formalism of GP regression is first reviewed, particularly in relation to its application in learning local atomic energies and forces. 
For accurate regression it is fundamental to incorporate prior knowledge into the GP kernel function. 
To this end, this chapter details how properties of smoothness, invariance and interaction order of a force field can be encoded into corresponding kernel properties.
A range of kernels is then proposed, possessing all the required properties and an adjustable parameter $n$ governing the interaction order modelled.
The order $n$ best suited to describe a given system can be found automatically within the Bayesian framework by maximisation of the marginal likelihood.
The procedure is first tested on a toy model of known interaction and later applied to two real materials described at the DFT level of accuracy.
The models automatically selected for the two materials were found to be in agreement with physical intuition. 
More in general, it was found that lower order (simpler) models should be chosen when the data are not sufficient to resolve more complex interactions.
Low $n$ GPs can be further sped up by orders of magnitude by constructing the corresponding tabulated force field, here named \q{MFF}.}

\section{Introduction}

\label{sec:introduction}

The No Free Lunch (NFL) theorems proven by D. H. Wolpert in 1996 state that no learning algorithm can be considered better than any other (and than random guessing) when its performance is averaged uniformly over all possible functions \cite{computation:1996fp}. 
Although functions appearing in real world problems are certainly not uniformly distributed, this remarkable result seems to suggest that the search for the \q{best}  machine learning (ML) algorithm able to learn any function in an \q{agnostic} fashion is groundless, and strongly justifies current efforts within the physics and chemistry communities aimed at the development of ML techniques that are particularly suited to tackle a {\em given} problem, for which prior knowledge is available and exploitable.  

In the context of machine learning force field (ML-FF) generation this resulted in a proliferation of different approaches based on artificial neural networks (NN) \cite{Skinner:1995ea,Behler:2007fe,Kondor:2018ub,Gastegger:2015dl,Manzhos:2014jb,Geiger:2013ea,Kuritz:2018wp,Schutt:2017kd,Lubbers:2018dv,Schutt:2018gl}, Gaussian process (GP) regression \cite{Bartok:2010fj,Li:2015eb,Glielmo:2017dj,Glielmo:2018bm,Zeni:2018to,Szlachta:2014jh} or linear expansions on properly defined bases \cite{Thompson:2015dw,Shapeev:2016kna,Takahashi:2017th}.
Particularly within GP regression (the method predominantly discussed in this chapter), a considerable effort was directed towards the inclusion of the known physical symmetries of the target system (translations, rotations and permutations) in the algorithm as a prior piece of information.
Among these, rotation symmetry proved the most cumbersome one to deal with, and received special attention. 
This typically involved either building  explicitly invariant descriptors (as the Li et al. feature-matrix based on internal vectors \cite{Li:2015eb}), or imposing the symmetry via an invariant \cite{Bartok:2013cs} or covariant \cite{Glielmo:2017dj} integral to learn energies or forces.
Clearly, many more detailed recipes than those featuring in the list above would be possible in virtually all situations, 
making the problem of selecting a single model for a particular task both interesting and unavoidable.   
In the following, we will argue that a good way of choosing among competing explanations is to follow the long-standing Occam's razor principle and select the simplest model that is still able to provide a satisfactory explanation~\cite{Jefferys:1992kl,Rasmussen:vk,Ghahramani:2015ee}. 

This general idea has found rigorous mathematical formulations. Within statistical learning theory, the complexity of a model can be measured by calculating its Vapnik Chervonenkis (VC) dimension \cite{Vapnik:2015bz,Vapnik:1998uq}.
The \emph{VC dimension} of a model then relates to its \emph{sample complexity} (i.e., the number of points needed to effectively train it) as one can prove that the latter is bounded by a monotonic function of the former \cite{Kearns:1994wq, Vapnik:1998uq}.
Similar considerations can also be made in a Bayesian context by noting that models with prior distributions concentrated around the true function (i.e. {\em simpler} models) have a lower sample complexity and will hence learn faster \cite{Theory:ub}.
\begin{figure}[t]
	\sidecaption[t]
	\includegraphics[width=0.6\columnwidth]{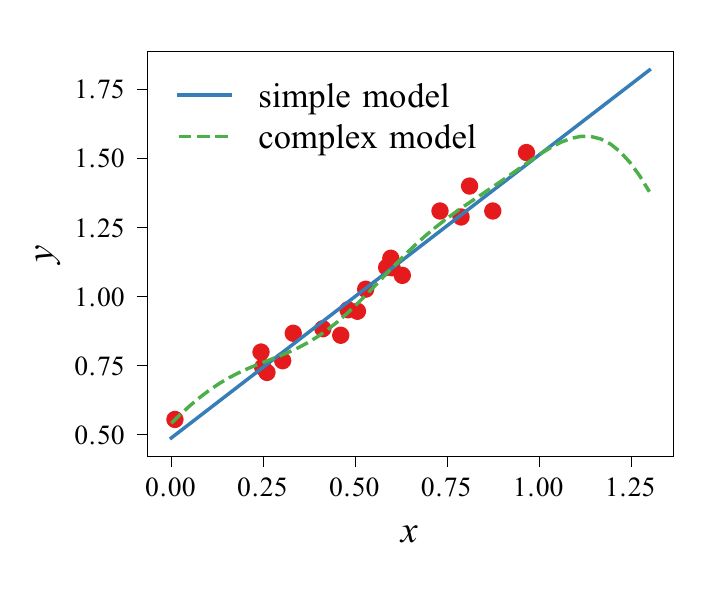}
	\caption{A simple linear model (blue solid line) and a complex GP model (green dashed line) are fitted to some data points. In this situation, if we have prior knowledge that a linear trend underpins the data, we should enforce the blue model {\em a priori}; otherwise we should select the blue model by Occam's razor after the data becomes available, since it is the simplest one. The advantages of this choice lie in the greater interpretability and extrapolation power of the simpler model.}
	\label{fig:1}      
\end{figure}
The above considerations suggest that a principled approach to learn a force field is to incorporate as much prior knowledge as is available on the function to be learned and the particular system at hand. 
When prior knowledge is not enough to decide among competing models, these should all be trained and tested, after which the simplest one that is still compatible with the desired target accuracy should be selected. 
This approach is illustrated in Figure~\ref{fig:1}, where two competing models are considered for a one dimensional data set.

In the rest of this chapter we provide a step-by-step guide to the incorporation of prior knowledge and to model selection in the context of Bayesian regression based on GP priors (Section \ref{sec:theory}) and show how these ideas can be applied in practice (Section \ref{sec:practice}). 
Section \ref{sec:theory} is structured as follows. In Section \ref{subsec:theory_1} we give a pedagogical introduction to GP regression, with a focus on the problem of learning a local energy function.
In Section \ref{subsec:theory_local_energies_from_forces} we show how a local energy function can be learned in practice when using a database containing solely total energies and/or forces.
In Section \ref{subsec:theory_prior} we then review the ways in which physical prior information can (and should) be incorporated in GP kernel functions, focusing on smoothness (\ref{subsubsec:theory_priora}), symmetries (\ref{subsubsec:theory_priorb}) and interaction order (\ref{subsubsec:theory_priorc}). 
In Section \ref{subsec:theory_building_nkernels} we make use of the preceding section's results to define a set of kernels of tunable complexity that incorporate as much prior knowledge as is available on the target physical system.   
In Section \ref{subsec:theory_model_selection} we show how Bayesian model selection provides a principled and \q{automatic} choice of the simplest model suitable to describe the system.
For simplicity, throughout this chapter only systems of a single chemical species are discussed, but in Section \ref{subsec:theory_multispecies_kernels} we briefly show how the ideas presented can be straightforwardly extended to model multispecies systems.

Section \ref{sec:practice} focuses on the practical application of the ideas presented. 
In particular, Section \ref{subsec:practice_1} describes an application of the model selection method described in Section \ref{subsec:theory_model_selection} to two different Nickel environments, represented as different subsets of a general Nickel database.  
We then compare the results obtained from this Bayesian model selection technique with those provided by a more heuristic model selection approach and show how the two methods, while being substantially different and optimal in different circumstances, typically yield similar results.
The final Section \ref{subsec:practice_2} discusses the computational efficiency of GP predictions, and explain how a very simple procedure can increase by several orders of magnitude the evaluation speed of certain classes of GPs when on-the-fly training is not needed. 
The code used to carry out such a procedure is freely available as part of the \q{MFF} Python package \cite{mff_package}.

\section{Nonparametric $n$-body force field construction}
\label{sec:theory}

The most straightforward well defined local property accessible to QM calculations is the force on atoms, which can be easily computed by way of the Hellman-Feynman theorem \cite{Feynman:1939bg}. 
Atomic forces can be machine learned directly in various ways, and the resulting model can be used to perform molecular dynamics simulations, probe the system's free energy landscape, etc.  \cite{Li:2015eb,Glielmo:2017dj,Zeni:2018to,Botu:2015kb,Kruglov:2017ju}. 
We can however also define a \emph{local energy} function $\epsilon(\rho)$ representing the energy $\epsilon$ of an atom given a representation $\rho$ of the set of positions of all the atoms surrounding it within a cutoff distance. 
Such a set of positions is typically called an \emph{atomic environment} or an \emph{atomic configuration}, and $\rho$ could simply be a list of the atomic species and positions expressed in Cartesian coordinates, or any suitably chosen representation of these \cite{Ferre:2015dq,Bartok:2013cs,Li:2015eb,Glielmo:2018bm}.

Although local energies are not well-defined in quantum calculations, in the following section we will be focusing on GP models for learning this somewhat accessory function $\epsilon(\rho)$, as this makes it easier to understand the key concepts \cite{Bartok:2015iw}.
We will also assume for simplicity that our ML model is trained on a database of local configurations and energies, although in practice $\epsilon(\rho)$ is machine-learned from the atomic forces and total energies produced by QM codes. 
The details of how this can be practically done will be discussed in Section~\ref{subsec:theory_local_energies_from_forces}.  

\subsection{Gaussian process regression}
\label{subsec:theory_1}

In order to learn the local energy function $\epsilon (\rho)$ yielding the energy of the atomic configuration $\rho$ we assume to have access to a database of reference calculations $\mathscr{D} = \{(\epsilon_i^r, \rho_i)\}_{i=1}^{N}$ composed by $N$ local atomic configurations $\boldsymbol{\rho} = (\rho_1,\dots, \rho_N)^T$ and their corresponding energies
$\boldsymbol{\epsilon}^r = (\epsilon_1^r,\dots, \epsilon_N^r)^T$. 
It is assumed that the energies have been obtained as 
\begin{equation}
\label{eq:database_creation}
\epsilon_i^r = \epsilon(\rho_i) + \xi_i
\end{equation}
where the noise variables $\xi_i$ are independent zero mean Gaussian random variables ($\xi_i \sim\mathscr{N}(0, \sigma_n^2)$). 
This noise in the data can be imagined to represent the combined uncertainty associated with both training data and model used. 
For example, an important source of uncertainty is the \emph{locality error} resulting from the assumption of a finite cutoff radius, outside of which atoms are treated as non-interacting. 
This assumption is necessary in order to define local energy functions but it never holds exactly. 

The power of GP regression lies in the fact that $\epsilon(\rho)$ is not constrained to be a given parametric functional form as in standard fitting approaches, but it is rather assumed to be distributed as a Gaussian stochastic process, typically with zero mean 
\begin{equation}
\epsilon(\rho) \sim \mathscr{GP}(0, k(\rho,\rho'))
\label{eq:prior_GP}
\end{equation}
where $k$ is the \emph{kernel function} of the GP (also called \emph{covariance function}). 
This notation signifies that for any finite set of input configurations $\boldsymbol{\rho}$, the corresponding set of local energies $\boldsymbol{\epsilon} =( \epsilon(\rho_1), \dots,\epsilon(\rho_N))^T$ will be distributed according to a multivariate Gaussian distribution whose covariance matrix is constructed through the kernel function:
\begin{equation}
\begin{cases}

p(\boldsymbol{\epsilon} \mid  \boldsymbol{\rho})  &= \mathscr{N}(\boldsymbol{0}, \boldsymbol{K}) \\
\boldsymbol{K}  			&= \begin{pmatrix} k(\rho_1,\rho_1) & \cdots & k(\rho_1,\rho_N)\\
																\vdots & \ddots & \vdots\\
																k(\rho_N,\rho_1) & \cdots & k(\rho_N,\rho_N)
\end{pmatrix}.
\end{cases}
\end{equation}
Given that both $\xi_i$ and $\epsilon(\rho_i)$ are normally distributed, and since the sum of two Gaussian random variables is also a Gaussian variable, one can write down the distribution of the reference energies $\epsilon^r_i$ of Eq.~(\ref{eq:database_creation}) as a new normal distribution whose covariant matrix is the sum of the original two:
\begin{equation}
\begin{cases}

p(\boldsymbol{\epsilon}^r \mid  \boldsymbol{\rho})  &= \mathscr{N}(\boldsymbol{0}, \boldsymbol{C}) \\
\boldsymbol{C}  			&= \boldsymbol{K} + \boldsymbol{1}\sigma_n^2.
\end{cases}
\label{eq:energy_prediction}
\end{equation}
Building on this closed form (Gaussian) expression for the probability of the reference data, we can next calculate the \emph{predictive distribution} i.e., the probability distribution of the local energy value $\epsilon^*$ associated with a new target configuration $\rho^*$, for the given training dataset 
$\mathscr{D}=(\boldsymbol{\rho},\boldsymbol{\epsilon}^r)$--the interested reader is referred to the two excellent references~\cite{Williams:2006vz, Bishop:998831} for details on the derivation. This is:
\begin{equation}
\begin{cases}
p(\epsilon^* \mid \rho^* , \mathscr{D})  &= \mathscr{N}(\hat{\epsilon}(\rho^*), \hat{\sigma}^2(\rho^*)) \\
\hat{\epsilon} (\rho^*)													  &= \boldsymbol{k}^T \boldsymbol{C}^{-1} \boldsymbol{\epsilon}^r \\
\hat{\sigma}^2 (\rho^*)															   	  &= k(\rho^*, \rho^*) + \sigma^2_n -  \boldsymbol{k}^T\boldsymbol{C}^{-1}\boldsymbol{k} 
\end{cases},
\label{eq: energy_pred}
\end{equation}
where we defined the vector $\boldsymbol{k}  = (k(\rho^*, \rho_1 ), \dots, k(\rho^*, \rho_N ))^T$ . 
The mean function $\hat{\epsilon}(\rho)$ of the predictive distribution is now our \q{best guess} for the true underlying function as it can be shown that it minimises expected error
\footnote{
Choosing a squared error function $L=(\bar{\epsilon}(\rho)- \epsilon)^2$, the expected error under the posterior distribution reads
$
\langle L \rangle = \int d\epsilon \, p(\epsilon \mid \rho, \mathscr{D}) (\bar{\epsilon}(\rho)- \epsilon)^2.
$
%
%
Minimising this quantity with respect to the unknown optimal prediction $\bar{\epsilon}(\rho)$ can be done by equating the functional derivative $\delta \langle L \rangle / \delta \bar{\epsilon}(\rho)$ to zero, yielding the condition
$
(\bar{\epsilon}(\rho) - \langle \epsilon \rangle ) = 0,
$
proving that the optimal estimate corresponds to the mean $\hat{\epsilon}(\rho)$ of the predictive distribution in Eq.~(\ref{eq: energy_pred}).
One can show that choosing an absolute error function $L=\left| \bar{\epsilon}(\rho)- \epsilon \right|$ makes the mode of the predictive distribution the optimal estimate, this however coincides with the mean in the case of Gaussian distributions.}.

The mean function is often equivalently written down as a linear combination of kernel functions evaluated over all database entries
\begin{equation}
\hat{\epsilon}(\rho) = \sum_{d=1}^N  k(\rho,\rho_d)\alpha_d,
\label{eq:explicit_GP_pred}
\end{equation} 
where the coefficients are readily computed as $\alpha_d = (\boldsymbol{C}^{-1} \boldsymbol{\epsilon})_d $.
The posterior variance of $\epsilon^*$ 
provides a measure of the uncertainty associated with the prediction, normally expressed as the standard deviation $\hat{\sigma}(\rho)$.

\begin{figure}[t]
	\begin{centering}
			\subfloat[\label{fig:prior}]{\includegraphics[width=0.5\columnwidth]{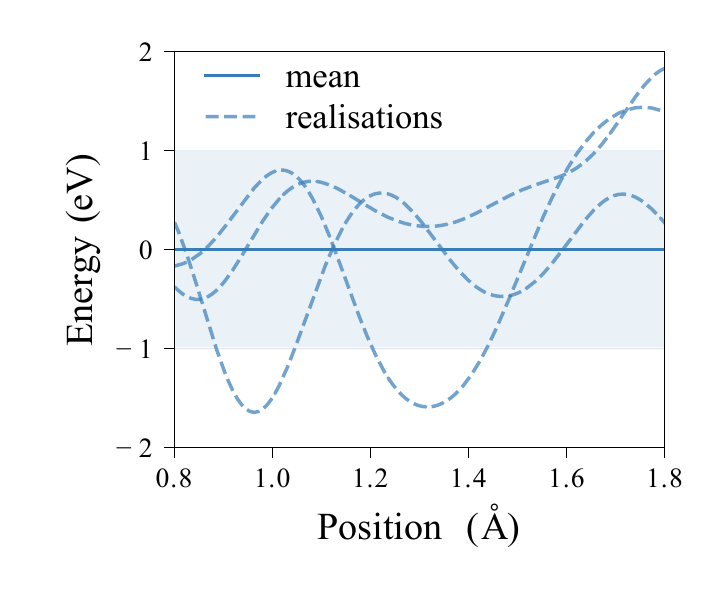}
				
			}\subfloat[\label{fig:posterior}]{\includegraphics[width=0.5\columnwidth]{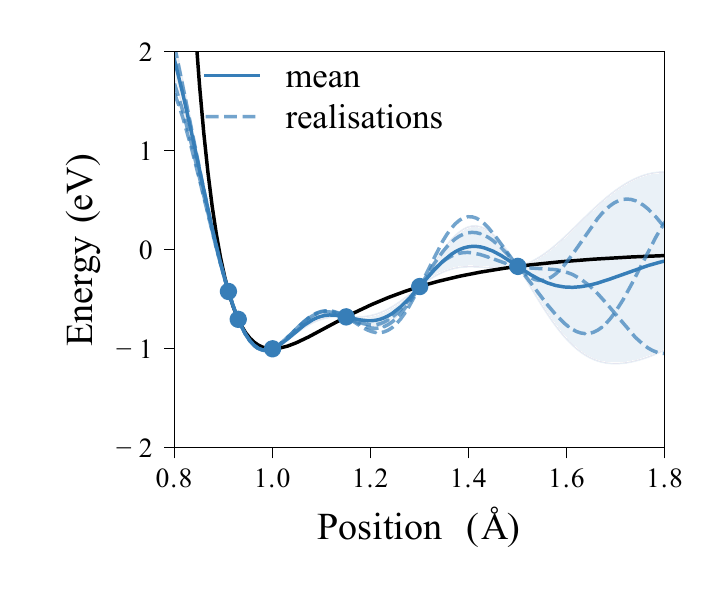}
			}
		\par\end{centering}
		
	\caption{Pictorial view of GP learning of a LJ dimer. Panel (a): mean, standard deviation and random realisations of the prior stochastic process, which represents our belief on the dimer interaction before any data is seen. Panel (b): posterior process, whose mean passes through the training data and whose variance provides a measure of uncertainty.}
	\label{fig:GP}
\end{figure}

The GP learning process can be thought of as an update of the prior distribution Eq.~(\ref{eq:prior_GP}) into the posterior Eq.~(\ref{eq: energy_pred}).
This update is illustrated in Figure \ref{fig:GP}, in which GP regression is used to learn a simple Lennard Jones (LJ) profile from a few dimer data.
In particular, Figure \ref{fig:prior} shows the prior GP (Eq.~(\ref{eq:prior_GP}) while Figure \ref{fig:posterior} shows the posterior GP, whose mean and variance are those of the predictive distribution Eq.~(\ref{eq: energy_pred}).
By comparing the two panels one notices that the mean function (equal to zero in the prior process) approximates the true function (black solid line) by passing through the reference calculations. 
Clearly, the posterior standard deviation (uniform in the prior) shrinks to zero at the points where data is available (as we set the intrinsic noise $\sigma_n$ to zero) to then increase again away from them.
Three random function samples are also shown for both prior and posterior process.

\subsection{Local energy from global energies and forces}
\label{subsec:theory_local_energies_from_forces}

The forces acting on atoms are well defined local property accessible to QM calculations, easily computed by way of the Hellman-Feynman theorem \cite{Feynman:1939bg}.
As a consequence,  GP regression can in principle be used to learn a force field directly on a database of quantum forces, as done for instance in Refs.~\cite{Glielmo:2017dj,Li:2015eb,Botu:2015kb}. 
Local atomic energies on the contrary cannot be computed in QM calculations, which can only provide the \emph{total} energy of the full system.
However, the material presented in the previous section, in addition to being of pedagogical importance, is still useful in practice since local energy functions can be learned from observations of total energies and forces only.

Mathematically this is possible since any sum, or derivative, of a Gaussian process is also a Gaussian process \cite{Williams:2006vz}, and the main ingredients needed for learning are hence the covariances (kernels) between these Gaussian variables.
In the following, we will see how kernels for total energies and forces can be obtained starting from a kernel for local energies, and how these derived kernels can be used to learn a local energy function from global energy and force information.

\runinhead{Total energy kernels.}
The total energy of a system can be modelled as a sum of the local energies associated to each local atomic environment
\begin{equation}
E( \{ \rho_a \}) = \sum_{a=1}^{N_a} \varepsilon(\rho_a)
\end{equation}
and if the local energy functions $\varepsilon$ in the above equation are distributed according to a zero mean GP, then also the global energy $E$ will be GP variable with zero mean. 
To calculate the kernel functions $k^{\varepsilon E}$ and $k^{EE}$ providing the covariance between local and global energies and between two global energies one simply needs to take the expectation with respect to the GP of the corresponding products

\noindent\begin{minipage}{.5\linewidth}
\begin{equation}
\begin{split}
k^{\varepsilon E}(\rho_a, \{\rho_b' \}) &= \langle \varepsilon( \rho_a ) E( \{\rho_b' \}) \rangle \\
& = \sum_{b=1}^{N'_a}  \langle \varepsilon(\rho_a) \varepsilon(\rho_b') \rangle \\
& = \sum_{b=1}^{N'_a} k(\rho_a, \rho_b).
\end{split}
\end{equation}
\end{minipage}%
\begin{minipage}{.5\linewidth}
\begin{equation}
\begin{split}
k^{EE}(\{\rho_a \}, \{\rho_b' \}) &= \langle E( \{\rho_a \}) E( \{\rho_b' \}) \rangle \\
& = \sum_{a=1}^{N_a} \sum_{b=1}^{N'_a}  \langle \varepsilon(\rho_a) \varepsilon(\rho_b') \rangle \\
& =  \sum_{a=1}^{N_a} \sum_{b=1}^{N'_a} k(\rho_a, \rho_b).
\end{split}
\end{equation}
\end{minipage}

\noindent
Note that we have allowed the two systems to have a different number of particles $N_a$ and $N'_a$ and that the final covariance functions can be entirely expressed in terms of local energy kernel functions $k$.

\vspace{5 pt}

\runinhead{Force kernels.}

The force $\mathbf{f}(\{ \rho_a \}^p)$ on an atom $p$ at position $\mathbf{r}_p$ is defined as the derivative
\begin{equation}
\mathbf{f}(\{ \rho_a \}^p) = - \frac{\partial E( \{ \rho_a \}^p)}{\partial \mathbf{r}_p},
\end{equation}
where by virtue of the existence of a finite cutoff radius of interaction, only the set of configurations $\{ \rho_a \}^p$ that contain atom $p$ within their cutoff function contribute to the force on $p$.
This quantity is also a GP \cite{Williams:2006vz} and the corresponding kernels between forces and between forces and local energies can be easily obtained by differentiation as described in Refs.~\cite{Williams:2006vz,Macedo:2008vq}. They read

\noindent
\begin{minipage}{.5\linewidth}
\begin{equation}
\mathbf{k}^{\varepsilon \mathbf{f}}(\rho_a , \{ \rho_b \}^p) 
= - \sum_{\{ \rho_b \}^q} \frac{\partial k(\rho_a, \rho_b) }{\partial \mathbf{r}_q^{\rm{T}}}
\end{equation}
\end{minipage}%
\begin{minipage}{.5\linewidth}
\begin{equation}
\mathbf{K}^{\mathbf{f}\mathbf{f}}(\{ \rho_a \}^p, \{ \rho_b \}^q) 
= \sum_{\{ \rho_a \}^p} \sum_{\{ \rho_b \}^q} \frac{\partial^2 k_n(\rho_a, \rho_b) }{\partial \mathbf{r}_p \partial \mathbf{r}_q^{\rm{T}} }
\end{equation}
\end{minipage}.

\runinhead{Total energy-force kernel.}

Learning from both energies and forces simultaneously is also possible. 
One just needs to calculate the extra kernel $\mathbf{k}^{\mathbf{f}E}$ comparing the two quantities in the database
\begin{equation}
\mathbf{k}^{\mathbf{f}E}(\{ \rho_a \}^p, \{\rho_b' \}) = - \sum_{\{ \rho_a \}^p} \sum_{b=1}^{N'} \frac{\partial k(\rho_a, \rho_b)}{\partial \mathbf{r}_p}.
\end{equation}
%
\vspace{1cm}

To clarify how the kernels described above can be used in practice, it is instructive to look at a simple example. 
Imagine having a database made up of a single snapshot coming from an \emph{ab initio} molecular dynamics of $N$ atoms, hence containing a single energy calculation and $N$ forces.
Learning using these quantities  would involve building a $N +1 \times N +1$ block matrix $\mathbb{K}$ containing the covariance between every pair
\begin{equation}
\mathbb{K}=\begin{pmatrix}
k^{EE}(\{\rho_a\},\{\rho_b\}) & \mathbf{k}^{E\mathbf{f}}(\{\rho_a\},\{\rho_b\}^1) & \cdots & \mathbf{k}^{E\mathbf{f}}(\{\rho_a\},\{\rho_b\}^N)\\
\mathbf{k}^{\mathbf{f}E}(\{\rho_a\}^1,\{\rho_b\}) & \mathbf{K}^{\mathbf{ff}}(\{\rho_a\}^1,\{\rho_b\}^1) & \cdots & \mathbf{K}^{\mathbf{ff}}(\{\rho_a\}^1,\{\rho_b\}^N)\\
\vdots & \vdots & \ddots & \vdots \\
\mathbf{k}^{\mathbf{f}E}(\{\rho_a\}^N,\{\rho_b\}) & \mathbf{K}^{\mathbf{ff}}(\{\rho_a\}^N,\{\rho_b\}^1)  & \cdots & \mathbf{K}^{\mathbf{ff}}(\{\rho_a\}^N,\{\rho_b\}^N)
\end{pmatrix}.
\label{eq:big_gram_matrix}
\end{equation}
As is clear from the above equation, each block is either a scalar (the energy-energy kernel in the top left), a $3\times 3$ matrix (the force-force kernels) or a vector (the energy-force kernels). 
The full dimension of $\mathbb{K}$ is hence $(3N+1)\times (3N+1)$.

Once such a matrix is built and the inverse $ \mathbb{C}^{-1} = [\mathbb{K} + \mathbb{I}\sigma_n^2]^{-1}$ computed, the predictive distribution for the value of the latent local energy variable can be easily written down.
For notational convenience, it is useful to define the vector $\{ x_i \}_{i=1}^{N}$ containing all the quantities in the training database and the vector $\{ t_i \}_{i=1}^{N}$ specifying their type (meaning that $t_i$ is either $E$ or $\mathbf{f}$ depending on the type of data point contained in $x_i$).
With this convention the predictive distribution for the local energy takes the form
\begin{equation}
	\begin{split}
	p(\varepsilon^* \mid \rho^*, \mathscr{D}) & = \mathscr{N}(\hat{\varepsilon}(\rho^*), \hat{\sigma}^2(\rho^*) ) \\
	\hat{\varepsilon}(\rho^*) & = \sum_{ij} k^{\varepsilon t_i}(\rho^*, \rho_i) \mathbb{C}^{-1}_{ij} x_j \\
	\hat{\sigma}^2(\rho^*) & = k(\rho^*, \rho^*) - \sum_{ij} k^{\varepsilon t_i}(\rho^*, \rho_i) \mathbb{C}^{-1}_{ij} k^{t_j \varepsilon }(\rho_j, \rho^*),\\
	\end{split}
\end{equation}
where the products between $x_j$, $\mathbb{C}^{-1}_{ij}$ and $k^{t_j \varepsilon }$ are intended to be between scalars, vectors or matrices depending on the nature of the quantities involved.

\vspace{1cm}

\subsection{Incorporating prior information in the kernel}
\label{subsec:theory_prior}

Choosing a Gaussian stochastic process as prior distribution over the local energies $\epsilon (\rho)$ rather than a parametrised functional form brings a few key advantages. 
A much sought advantage is that it allows grater flexibility: one can show that in general a GP corresponds to a model with an infinite number of parameters, and with a suitable kernel choice can act as a \q{universal approximator}: capable of learning any function if provided with sufficient training data \cite{Williams:2006vz}. 
A second one is a greater ease of design: the kernel function must encode all prior information about the local energy function, but typically contains very few free parameters (called \emph{hyperparameters}) which can be tuned, and such tuning is typically straightforward.   
Third, GPs offer a coherent framework to predict the uncertainty associated with the predicted quantities via the posterior covariance. 
This is typically not possible for classical parametrised $n$-body force fields.   

All this said, the high flexibility associated with GPs could easily become a drawback when examined from the point of view of computational efficiency. 
Broadly, it turns out that for maximal efficiency (which takes into account both accuracy and speed of learning and prediction) one should constrain this flexibility in physically motivated ways, essentially by incorporating prior information in the kernel.
This will reduce the dimensionality of the problem e.g., by choosing to learn energy functions of significantly fewer
variables than those featuring in the configuration $\rho$ ($3N$ for $N$ atoms). 

To effectively incorporate prior knowledge into the GP kernel it is fundamental to know the relation between important properties of the modelled energy and the corresponding kernel properties.
These are presented in the remainder of this section for the case of local energy kernels. 
Properties of smoothness, invariance to physical symmetries, and interaction order are discussed in turn.

\subsubsection{Function smoothness}
\label{subsubsec:theory_priora}

\begin{center}
	\begin{figure*}
		\begin{centering}
			\subfloat[\label{fig:smoothness_1}]{\includegraphics[width=0.5\columnwidth]{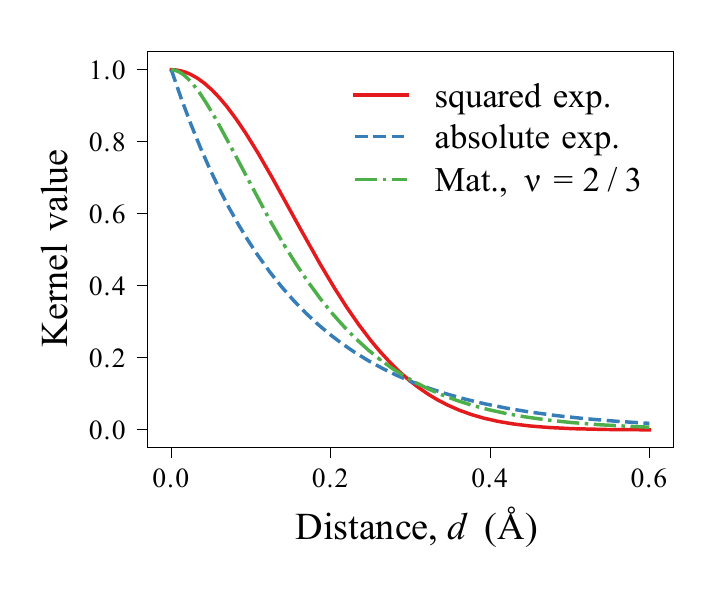}
				
			}\subfloat[\label{fig:smoothness_2}]{\includegraphics[width=0.5\columnwidth]{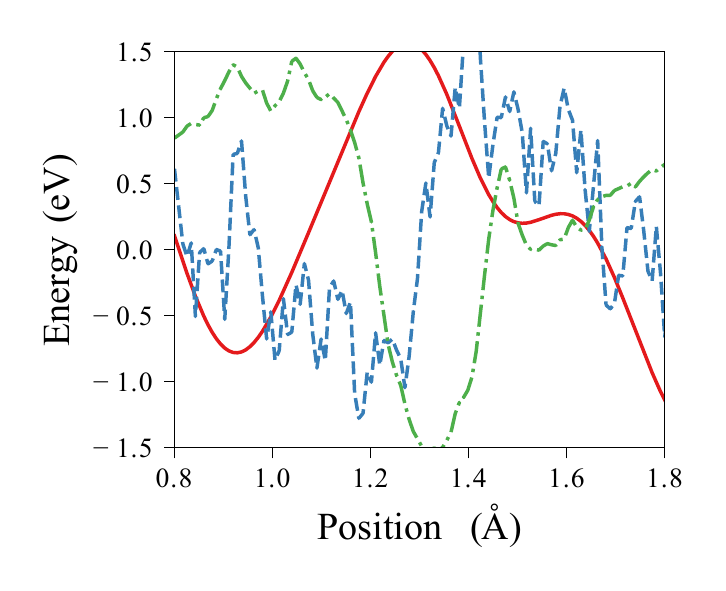}
			}
			\par\end{centering}
		
		\caption{Effect of three kernel functions on the smoothness of the corresponding stochastic processes.}
	\label{fig:smoothness}
	\end{figure*}
	\par\end{center}

The relation between a given kernel and the smoothness of the random functions described by the corresponding Gaussian stochastic process has been explored in detail \cite{Williams:2006vz,Bishop:998831}.
Kernels defining functions of arbitrary differentiability have been developed. 
For example, on opposite ends we find the so called \emph{squared exponential} ($k_{SE}$) and \emph{absolute exponential} ($k_{AE}$) kernels, defining respectively infinitely differentiable and nowhere differentiable functions:

\noindent
\begin{minipage}{.5\linewidth}
\begin{equation}
k_{SE}(d) = e^{-d^2/2\ell^2}
\end{equation}
\end{minipage}%
\begin{minipage}{.5\linewidth}
\begin{equation}
k_{AE}(d) = e^{-d/\ell}
\end{equation}
\end{minipage},

\vspace{0.1cm}
\noindent
where the letter $d$ represents the distance between two points in the metric space associated with the function to be learned (e.g., a local energy). 
The \emph{Mat\'{e}rn} kernel~\cite{Williams:2006vz,Bishop:998831} is a generalisation of the above mentioned kernels and allows to impose an arbitrary degree of differentiability depending on a parameter $\nu$:
\begin{equation}
	k_{M,\nu}(d) = \frac{2^{1-\nu}}{\Gamma(\nu)} \left( \sqrt{2 \nu} \frac{d}{\ell} \right)^{\nu} K_{\nu} \left( \sqrt{2 \nu} \frac{d}{\ell} \right),
\end{equation}
where $\Gamma$ is the gamma function and $K_{\nu}$ is a modified Bessel function of the second kind.

The relation between kernels and modelled function differentiability is illustrated by Figure \ref{fig:smoothness}, showing the three kernels mentioned above (Figure \ref{fig:smoothness_1}) along with typical samples from the corresponding GP priors (Figure \ref{fig:smoothness_2}).
The absolute exponential kernel has been found useful to learn atomisation energy of molecules \cite{Rupp:2012kxa,Rupp:2015et,Hansen:2013dp}, especially in conjunction with the discontinuous Coulomb matrix descriptor \cite{Rupp:2012kxa}. 
In the context of modelling useful machine learning force fields, a relatively smooth energy or force function is typically sought.
For this reason, the absolute exponential is not appropriate and has never been used while the flexibility of the Mat\'{e}rn covariance has only found limited applicability \cite{Chmiela:2017ff}.
In fact, the squared exponential has been almost always preferred, in conjunction with suitable representations $\rho$ of the atomic environment, \cite{Glielmo:2017dj,Zeni:2018to,Botu:2015kb,Deringer:2017ea}, and will be used also in this work. 

\subsubsection{Physical symmetries}
\label{subsubsec:theory_priorb}
Any energy or force function has to respect the symmetry properties listed below.

\runinhead{Translations.}

Physical systems are invariant upon rigid translations of all their components.
This basic property is relatively easy to enforce in any learning algorithm via a local representation of the atomic environments.
In particular, it is customary to express a given local atomic environment as the unordered set of $M$ vectors $\{\mathbf{r}_i\}_{i=1}^M$ going from the \q{central} atom to every neighbour lying within a given cutoff radius \cite{Glielmo:2017dj,Glielmo:2018bm,Bartok:2013cs,Ferre:2015dq}.
It is clear that any representation $\rho$ and any function learned within this space will be invariant upon translations.

\runinhead{Permutations.}

Atoms of the same chemical species are indistinguishable, and any permutation $\mathscr{P}$ of identical atoms in a configuration necessarily leaves energy (as well as the force) invariant. 
Formally one can write $ \epsilon(\mathscr{P} \rho) = \epsilon(  \rho) \, \forall \mathscr{P}$. 
This property corresponds to the kernel invariance 
\begin{equation}
k(\mathscr{P} \rho, \mathscr{P'} \rho') = k(\rho,\rho')  \,\,\,\, \forall \mathscr{P},\mathscr{P'}.
\label{eq:kernel_perm_inv}
\end{equation}
Typically, the above equality has been enforced either by the use of invariant descriptors \cite{Li:2015eb,Deringer:2017ea,Glielmo:2017dj, huo2017unified} or via an explicit invariant summation of the kernel over the permutation group \cite{Glielmo:2018bm, Zeni:2018to, Bartok:2013gf}, with the latter choice being feasible only when the symmetrisation involves a small number of atoms.

\runinhead{Rotations.}

The potential energy associated to a configuration should not change upon any rigid rotation $\mathscr{R}$ of the same (i.e., formally, $ \epsilon(\mathscr{R} \rho) = \epsilon( \rho) \, \forall \mathscr{R})$. 
Similarly to permutation symmetry, this invariance is expressed via the kernel property
\begin{equation}
k(\mathscr{R}\rho  ,\mathscr{R'} \rho') = k(\rho, \rho')  \,\,\,\, \forall \mathscr{R},\mathscr{R'}.
\label{eq:kernel_rot_inv}
\end{equation}
The use of rotation invariant descriptors to construct the representation $\rho$ immediately guarantees the above. 
Typical examples of such descriptors are the symmetry functions originally proposed in the context of neural networks \cite{Behler:2007fe,Behler:2011it}, the internal vector matrix \cite{Li:2015eb}, or the set of distances between groups of atoms \cite{Glielmo:2018bm,Deringer:2017ea, huo2017unified}.

Alternatively, a \q{base} kernel $k_b$ can be made invariant with respect to the rotation group via the following  symmetrisation (\q{Haar integral} over the full 3D rotation group): 
\begin{equation}
k( \rho,\rho')  = \int d\mathscr{R} \, k_b(\rho, \mathscr{R}\rho'). 
\label{eq:invariant_integral}
\end{equation}
Such a procedure (called \q{transformation integration} in the ML community \cite{Haasdonk:2007ff}) was first used to build a potential energy kernel in Ref. \cite{Bartok:2013cs}.

Learning forces, as well as other tensorial physical quantities (e.g., a stress tensor, or the (hyper)polarisability of a molecule), the learnt function must be \emph{covariant} under rotations.
This property can be formally written as $\mathbf{f}(\mathscr{R}\rho) = \mathbf{R}\mathbf{f} (\rho) \, \forall \mathscr{R}$ and, as shown in \cite{Glielmo:2017dj}, it translates at the kernel level to 
\begin{equation}
\mathbf{K}(\mathscr{R} \rho, \mathscr{R'} \rho')  = \mathbf{R} \mathbf{K}(\rho, \rho') \mathbf{R}^{\prime T}.
\end{equation}
Note that, since forces are three dimensional vectorial quantities, the corresponding kernels are $3 \times 3$ matrices \cite{Glielmo:2017dj,Micchelli:2004uv,Micchelli:2005ij}, here denoted by $\mathbf{K}$.

Designing suitable covariant descriptors is arguably harder than finding invariant ones.
For this reason, the automatic procedure proposed in Ref.~\cite{Glielmo:2017dj} to build covariant descriptors can be particularly useful.
Covariant matrix valued kernels are generated starting with an (easy to construct) scalar base kernel $k_b$ through a \q{covariant integral} 
\begin{equation}
\mathbf{K}( \rho,\rho')  = \int d\mathscr{R} \, \mathbf{R} k_b(\rho, \mathscr{R}\rho'). 
\label{eq:covariant_integral}
\end{equation}
This approach has been extended  to learn higher order tensors in Refs.~\cite{Bereau:2017vq,Grisafi:2017tw}.

\begin{figure}[t]
	\sidecaption[t]
	\includegraphics[width=0.5\columnwidth]{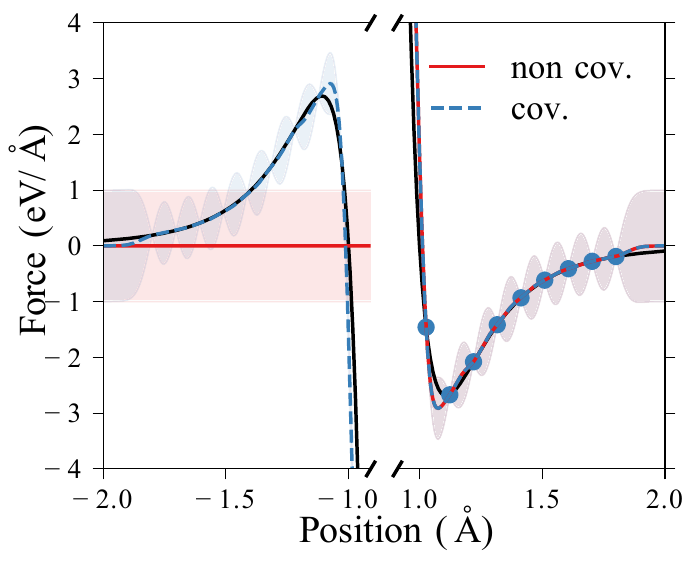}
	\caption{Learning the force profile of a 1D LJ dimer using data (blue circle) coming from one atom only. It is seen that a non covariant GP (solid red line) does not learn the symmetrically equivalent force acting on the other atom and it thus predict a zero force and maximum error. If covariance is imposed to the kernel via Eq.~(\ref{eq:covariant_integral}) (dashed blue line), then the correct equivalent (inverted) profile is recovered. Shaded regions represent the predicted 1sigma interval in the two cases.
	}
	\label{fig:covariance}
\end{figure}

Using rotational symmetry crucially improves the efficiency of the learned model. 
A very simple illustrative example of the importance of rotational symmetry is shown in Figure \ref{fig:covariance}, addressing an atomic dimer in which force predictions coming from a non-covariant squared exponential kernel and its covariant counterpart (obtained using Eq.~(\ref{eq:covariant_integral})) are compared. 
The figure reports the forces predicted to act on an atom, as a function of the position on the $x$-axis of the other atom, relative to the first.
So that, for positive $x$ values the figure reports the forces on the left atom as a function of the position of the right atom, while negative $x$ values will be associated to forces acting on the right atom as a function of the position of the left atom. 
In the absence of the covariance force properties, training the model on a sample of nine forces acting on the left atom, will populate correctly only the right side of the graph: a null force will be predicted to act on the right atom (solid red line on the left panel). 
However, the covariant transformation (in 1D, just a change of sign) will allow the transposition of the force field learned from one environment to the other, and thus the correct prediction of the (inverted) force profile
in the left panel. 

\subsubsection{Interaction order}
\label{subsubsec:theory_priorc}

Classical parametrised force fields are sometimes expressed as a truncated series of energy contributions of progressively higher $n$-body \q{interaction orders} \cite{Reddy:2016ic,Cisneros:2016hi,Stillinger:1985zz,Tersoff:1988jt}.
The procedure is consistent with the intuition that, as long as the series converges rapidly, truncating the expansion reduces the amount of data necessary for the fitting, and enables a likely higher extrapolation power to unseen regions of configuration space.
The lowest truncation order compatible with the target precision threshold is, in general, system dependent, as it will typically depend on the nature of the chemical interatomic bonds within the system. 
For instance, metallic bonding in a close-packed crystalline system might be described surprisingly well by a pairwise potential, while covalent bonding yielding a zincblend structure can never be, and it will always require three-body interactions terms to be present \cite{Glielmo:2017dj,Glielmo:2018bm}.
Restricting the order of a machine learning force field has proven to be useful for both neural network \cite{Yao:2017hc} and Gaussian process regression \cite{Glielmo:2017dj,Deringer:2017ea}.
In the particular context of GP-based ML-FFs, prior knowledge on the interaction order needs to be included in the form of an $n$-body kernel functions. 
A detailed and comprehensive exposition on how to do so was given in Ref.~\cite{Glielmo:2018bm}, and it will be summarised below and in the next subsection. 
The order of a kernel $k_n$ can be defined as the smallest integer $n$ for which the following property holds true

\begin{equation}
\frac{\partial^{n}k_{n}(\rho,\rho')}{\partial\mathbf{r}_{i_1}\cdots\partial\mathbf{r}_{i_n}}=0\label{eq: nBody_ker_def}
\hspace{1em} \hspace{1em}
\forall \, \,  \mathbf{r}_{i_1}\neq\mathbf{r}_{i_2}\neq\dots\neq\mathbf{r}_{i_n},
\end{equation}
where $\mathbf{r}_{i_1},\dots,\mathbf{r}_{i_n}$ are the positions of any choice of a set of $n$ different surrounding atoms. 
By virtue of linearity, the predicted local energy in Eq.~(\ref{eq:explicit_GP_pred}) will also satisfy the same property if $k_n$ does. 
Thus, Eq. (\ref{eq: nBody_ker_def}) implies that the central atom in a local configuration interacts with up to $n - 1$ other atoms simultaneously, making the learned energy $n$-body.

\subsection{Smooth, symmetric kernels of finite order $n$}
\label{subsec:theory_building_nkernels}
In the previous subsection we saw how the fundamental physical symmetries of energy and forces translate into the realm of kernels. 
Here we show how to build $n$-body kernels that possess these properties.

We start by defining a smooth translation- and and permutation-invariant 2-body kernel by summing all the squared exponential kernels calculated on the distances between the relative positions in $\rho$ and those in $\rho^{\prime}$ \cite{Glielmo:2017dj,Glielmo:2018bm,Zeni:2018to}
\begin{equation}
k_{2}(\rho,\rho')  =\sum_{\substack{i\in\rho, 
		j\in\rho'
	}
}e^{-\|\mathbf{r}_{i}-\mathbf{r}_{j}'\|^{2}/2 \ell^{2}}.
\label{eq: k2}
\end{equation}
As shown in \cite{Glielmo:2018bm}, higher order kernels can be defined simply as integer powers of $k_2$
\begin{equation}
k_{n}(\rho,\rho')=k_{2}(\rho,\rho')^{n-1}\label{eq: kn}
\end{equation}
Note that, by building $n$-body kernels using Eq.~(\ref{eq: kn}), one can avoid the exponential cost of summing over all $n$-plets that a more na{\"i}ve kernel implementation would involve.
This  makes it possible to model any interaction order paying only the quadratic computational cost of computing the 2-body kernel in Eq.~(\ref{eq: k2}).

Furthermore, one can at this point write the squared exponential kernel on the natural distance $d^2(\rho,\rho')=k_{2}(\rho,\rho)+ k_{2}(\rho',\rho') - 2 k_{2}(\rho,\rho')$ induced by the (\q{scalar product}) $k_2$ as a formal many body expansion:
\begin{align}
k_{MB}(\rho,\rho') & = \mathrm{e}^{-d^2(\rho,\rho')/2\ell^{2}}\nonumber \\
&= \mathrm{e}^{\frac{-k_{2}(\rho,\rho)- k_{2}(\rho',\rho')}{ 2 \ell^{2}} } \left[ 1+\frac{1}{\ell^{2}}k_{2}+\frac{1}{2!\ell^{4}}k_{3}+\frac{1}{3!\ell^{6}}k_{4}+\dots \right].
\label{eq: k_SE}
\end{align}
So that, assuming a smooth underlying function, the completeness of the series and the \q{universal approximator} property of the squared exponential \cite{Williams:2006vz,Hornik:1993dh} can be immediately seen to imply one another. 

It is important to notice that the scalar kernels just defined are \emph{not} rotation symmetric i.e., they do not respect the invariance property of  Eq.~(\ref{eq:kernel_rot_inv}).
This is due to the fact that the vectors $\mathbf{r}_{i}$ and $\mathbf{r}_{j}'$ featuring in Eq.~(\ref{eq: k2}) depend on the arbitrary reference frames with respect to which they are expressed. 
A possible solution would be given by carrying out the explicit symmetrisations provided by Eq.~(\ref{eq:invariant_integral}) (or  Eq.~(\ref{eq:covariant_integral}) if the intent is to build a force kernel).
The invariant integration Eq.~(\ref{eq:invariant_integral}) of $k_3$ is for instance a step in the construction of the (many-body) SOAP kernel \cite{Bartok:2013cs}, while an analytical formula for $k_n$ (with arbitrary $n$) has been recently proposed \cite{Glielmo:2018bm}. 
The covariant integral (Eq.~(\ref{eq:covariant_integral})) of finite-$n$ kernels was also successfully carried out (see Ref. \cite{Glielmo:2017dj}, which in particular contains a closed form expression for the $n=2$ matrix valued two-body force kernel).  

However, explicit symmetrisation via Haar integration invariably implies the evaluation of computationally expensive functions of the atomic positions. 
Motivated by this fact, one could take a different route and consider symmetric $n$-kernels defined, for any $n$, as functions of the effective rotation-invariant degrees of freedom of $n$-plets of atoms \cite{Glielmo:2018bm}.\
For $n=2$ and $n=3$ we can choose these degrees of freedom to be simply the interparticle distances occurring in  atomic pairs and triplets (other equally simple choices are possible, and have been used before, see Ref.~\cite{Deringer:2017ea}).  
The resulting kernels read:
\begin{align}
\ k_{2}^s(\rho,\rho') & =\sum_{\substack{i\in\rho\\
		j\in\rho'
	}
}\mathrm{e}^{-(r_{i}-r_{j}')^2/2\ell^2}\label{eq:2bodyk},\\
k_{3}^s(\rho,\rho') & =\sum_{\substack{i_1 > i_2\in\rho\\
		j_1 > j_2\in\rho'}} \,
	\sum_{\mathbf{P} \in \mathscr{P}} \mathrm{e}^{-\|(r_{i_1},r_{i_2},r_{i_1 i_2})^{\rm{T}}-\mathbf{P} (r_{j_1}',r_{j_2}',r_{j_1j_2}')^{\rm{T}} \|^2/2 \ell^2}. \label{eq:3bodyk}
\end{align}
where $r_i$ indicates the Euclidean norm of the relative position vector $\mathbf{r}_i$, and the sum over all permutations of 3 elements $\mathscr{P}$ ($\mid \mathscr{P} \mid = 6$) ensures the permutation invariance of the kernel (see Eq.~\eqref{eq:kernel_perm_inv}).

It was argued (and numerically tested) in \cite{Glielmo:2018bm} that these \emph{direct} kernels are as accurate as the Haar-integrated ones, while their evaluation is very substantially faster.
However, as is clear from Eqs.~(\ref{eq:2bodyk},\ref{eq:3bodyk}), even the construction of directly symmetric kernels becomes unfeasible for large values of $n$, since the number 
of terms in the sums grows exponentially. 
On the other hand, it is still possible to use Eq.~\eqref{eq: kn} to increase the integer order of an already symmetric $n'-$body kernel by elevating it to an integer power. 
As detailed in \cite{Glielmo:2018bm}, raising an already symmetric \q{input} kernel of order $n'$ to a power $\zeta$ in general produces a symmetric \q{output} kernel 
\begin{equation}
k_n^{\neg u}(\rho,\rho') = k_{n'}^s(\rho,\rho')^\zeta
\label{eq:non_unique_kernel}
\end{equation}
of order $n = (n' - 1)\zeta + 1 $.
We can assume that the input kernel was built on the effective degrees of freedom of the $n'$ particles in an atomic $n'$-uplet (as is the case e.g., the 2 and 3-kernels in Eqs.~(\ref{eq:2bodyk},\ref{eq:3bodyk})). 
The number of these degrees of freedom is 
$(3n'-6)$ for $n'>2$ (or just 1 for $n'=2$). 
Under this assumption, the output $n$-body kernel will depend on $\zeta (3n' - 6)$ variables (or just $\zeta$ variables for $n'=2$)
It is straightforward to check that this number is always smaller then the total number of degrees of freedom of $n$ bodies (here, $3n-6 = 3(n' - 1)\zeta -3$).
As a consequence, a rotation-symmetric kernel obtained as an integer power of an already  rotation-symmetric kernel will {\em not} be able to learn an {\em arbitrary} $n$-body interaction even if fully trained: its convergence predictions upon training on a given $n$-body reference potential will not be in general exact, and the prediction errors incurred will be specific to the input kernel and $\zeta$ exponent used.   
For this reason, kernels obtained via Eq.~\eqref{eq:non_unique_kernel} were defined \emph{non-unique} in Ref.~\cite{Glielmo:2018bm} (the superscript $\neg u$ in Eq. \eqref{eq:non_unique_kernel} stands for this). 

In practice, the non-unicity issue appears to be a severe problem only when the input kernel is a two-body kernel, and as such it depends only on the radial distances from the central atoms occurring in the two atomic configurations (cf. Eq. (\ref{eq:2bodyk})). 
In this case the non unique output $n$-body kernels will depend on $\zeta$-plets of radial distances, and will miss angular correlations encoded in the training data \cite{Glielmo:2018bm}.
On the contrary, a symmetric 3-body kernel (Eq. \eqref{eq:3bodyk}) contains angular information on all triplets in a configuration, and using this kernel as input will be able to capture higher interaction orders (as confirmed e.g., by the numerical tests performed in Ref.~\cite{Bartok:2013cs}).

Following the above reasoning, one can define a many-body kernel invariant over rotations as a squared exponential on the 3-body invariant distance $d^2_s(\rho,\rho')= k_3^s(\rho,\rho) + k_3^s(\rho',\rho') - 2 k_3^s(\rho,\rho')$, obtaining:
\begin{equation}
	k_{MB}^s(\rho,\rho')=\mathrm{e}^{-(k_3^s(\rho,\rho) + k_3^s(\rho',\rho') - 2 k_3^s(\rho,\rho'))/2\ell^2}.
	\label{eq:k_nu_SE}
\end{equation}
It is clear from the series expansion of the exponential function that this kernel is many-body in the sense of Eq.~(\ref{eq: nBody_ker_def}) and that the importance of high order contributions can be controlled by the hyperparameter $\ell$. 
With $\ell \ll 1$ high order interactions become dominant, while for $\ell \gg 1$ the kernel falls back to a 3-body description. 

For all values of $\ell$, the above kernel will however always encompass an implicit sum over all contributions (no matter how suppressed), being hence incapable of pruning away irrelevant ones even when a single interaction order is clearly dominant. 
Real materials often possess dominant interaction orders, and the ionic or covalent nature of their chemical bonding makes the many-body expansion converge rapidly. 
In these cases, an algorithm which automatically selects the dominant contributions, truncating this way the many-body series in Eq.~(\ref{eq: k_SE}), would represent an attractive option. This is the subject of the following section.

\subsection{Choosing the optimal kernel order}
\label{subsec:theory_model_selection}

In the previous sections we analysed how prior information can be encoded in the kernel function.
This brought us to designing kernels that implicitly define smooth potential energy surfaces and force fields with all the desired symmetries, and corresponding to a given interaction order (Eqs.~(\ref{eq:3bodyk}, \ref{eq:non_unique_kernel})). 
This naturally raises the problem of deciding the order $n$ best suited to describe a given system.  
A good conceptual framework for a principled choice is that of Bayesian model selection, which we now briefly review.

We start by assuming we are given a set of models $\{ \mathscr{M}_n^{\boldsymbol{\theta}} \}$ (each e.g., defined by a kernel function of given order $n$). 
Each model will be equipped with a vector of \emph{hyperparameters} $\boldsymbol{\theta}$,  (typically associated with the covariance lengthscale $\ell$, the data noise level $\sigma_n$, and similar). 
A fully Bayesian treatment would involve calculating the posterior probability of each candidate model, formally expressed via Bayes' theorem as
\begin{equation}
p( \mathscr{M}_n^{\boldsymbol{\theta}} \mid \boldsymbol{\rho} , \boldsymbol{\epsilon}^r ) 
= \frac
{p(\boldsymbol{\epsilon}^r\ \mid \boldsymbol{\rho}, \mathscr{M}_n^{\boldsymbol{\theta}} ) p(\mathscr{M}_n^{\boldsymbol{\theta}})}
{p(\boldsymbol{\epsilon}^r \mid \boldsymbol{\rho})},
\end{equation}
and selecting the model that maximises it.
However, often little \emph{a priori} information is available on the candidate models and their hyperparameters (or it is simply interesting to operate a selection unbiased by priors, and \q{let the data speak}). 
In such a case, the prior $p(\mathscr{M}_n^{\boldsymbol{\theta}})$ can be ignored as being flat and uninformative, and maximising the posterior becomes equivalent to maximising the \emph{marginal likelihood} $p(\boldsymbol{\epsilon}^r\ \mid \boldsymbol{\rho}, \mathscr{M}_n^{\boldsymbol{\theta}} )$ (here equivalent to the \emph{model evidence}.
\footnote{
The model evidence is conventionally defined as the integral over the hyperparameter space of the marginal likelihood times the hyperprior (cf.  \cite{Williams:2006vz}). We here simplify the analysis by jointly considering the model and its hyperparameters.
}
), and the optimal selection tuple $(n,{\boldsymbol{\theta}})$ can be hence chosen as
\begin{equation}
(\hat{n}, \hat{{\boldsymbol{\theta}}}) = \argmax_{(n, {\boldsymbol{\theta}})} \, p(\boldsymbol{\epsilon}^r\ \mid \boldsymbol{\rho}, \mathscr{M}_n^{\boldsymbol{\theta}}  ). 
\label{eq: ML}
\end{equation}
The marginal likelihood is an analytically computable normalised multivariate distribution, and it was given in Eq.~(\ref{eq:energy_prediction}).
\begin{figure}[t]
	\sidecaption[t]
	\includegraphics[width=0.55\columnwidth]{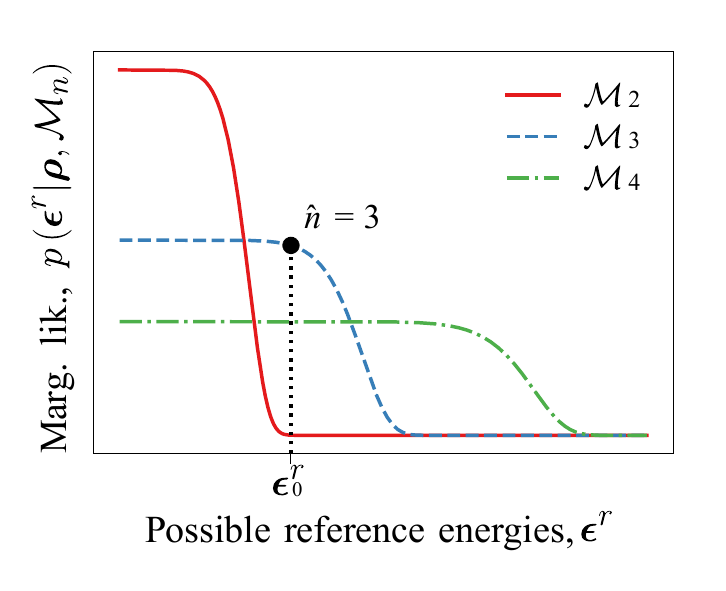}
	\caption{Cartoon of the marginal likelihood profile of three models of increasing complexity. More complex models can fit very different datasets $\boldsymbol{\epsilon}^r$, this is illustrated by the fact that their marginal likelihood is non-zero for a broader region of the dataset space (here pictorially one dimensional).}
	 \label{fig:model_selection}
\end{figure}

The maximisation in Eq.~(\ref{eq: ML}) can be thought of as a formalisation of the Occam's razor principle in our particular context. 
This is illustrated in Figure \ref{fig:model_selection}, which contains a cartoon of the marginal likelihood of three models of increasing complexity/flexibility (a useful analogy is to think of polynomials $P_n(x)$ of increasing order $n$, the likelihood representing how well these would fit a set of measurements $\epsilon^r$ of an unknown function $\epsilon(x)$). 
By definition, the most complex model in the figure is the green one, as it assigns a non-zero probability to the largest domain of possible outcomes, and would thus be able to explain the widest range of datasets.
Consistently, the simplest model is the red one, which is instead restricted to the smallest dataset range (in our analogy, a straight line will be able to fit well fewer data sets than a fourth order polynomial). 
Once a reference database $\epsilon_0^r$ is collected, it is immediately clear that the $\mathscr{M}_3$ model with highest likelihood $p(\boldsymbol{\epsilon} \mid \boldsymbol{\rho}, \mathscr{M}_n^{\boldsymbol{\theta}})$ at $\epsilon^r = \epsilon_0^r$
is the simplest that is still able to explain it (the blue one in Figure \ref{fig:model_selection}). 
Indeed, the even simpler model $\mathscr{M}_2$ is not likely to explain the data, the more complex model $\mathscr{M}_4$ can explain more than is necessary for compatibility with the $\epsilon_0^r$ data at hand, and thus produces a lower likelihood value, due to normalisation.  

To see how these ideas work in practice, we first test them on a simple system with controllable interaction order, while real materials are analysed in the next section. 
We here consider a one dimensional chain of atoms interacting via an \emph{ad hoc} potential of order $n^t$ ($t$ standing for \q{true}) 
\footnote{
The $n$-body toy model used was set up as a hierarchy of two body interactions defined via the negative Gaussian function
$
\epsilon^g(d) = - e^{-\frac{(d-1)^2}{2}}
$
This pairwise interaction, depending only on the distance $d$ between two particles, was then used to generate $n$-body local energies as 
$\epsilon_n(\rho) = \sum_{i_1\neq \dots \neq i_{n-1}} \epsilon^g(x_{i_1}) \epsilon^g(x_{i_{2}}-x_{i_{1}}) \dots \epsilon^g(x_{i_{n-2}}-x_{i_{n-1}})$
where $x_{i_1},\dots, x_{i_n-1}$ are the positions, relative to the central atom, of $n-1$ surrounding neighbours.}.

For each value of $n^t$, we generate a database of $N$ randomly sampled configurations and associated energies.
To test Bayesian model selection, for different reference $n^t$ and $N$ values and for fixed $\sigma_n \approx 0$ (noiseless data), we selected the optimal lengthscale parameter $\ell$ and interaction order $n$ of the $n$-kernel in Eq.~(\ref{eq: kn}) by solving the maximisation problem of Eq.~(\ref{eq: ML}). 
This procedure was repeated 10 times to obtain statistically significant conclusions, the results were however found to be very robust in the sense that they did not depend significantly on the specific realisation of the training dataset.
\begin{center}
	\begin{figure*}[h]
		\begin{centering}
			\subfloat[$n^t = 2$\label{fig:ms_3a}]{\includegraphics[width=0.33\columnwidth]{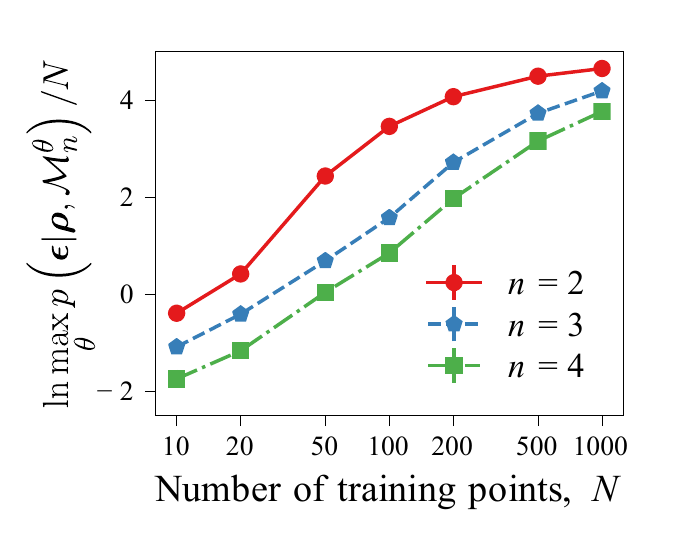}}
			\subfloat[$n^t = 3$\label{fig:ms_3b}]{\includegraphics[width=0.33\columnwidth]{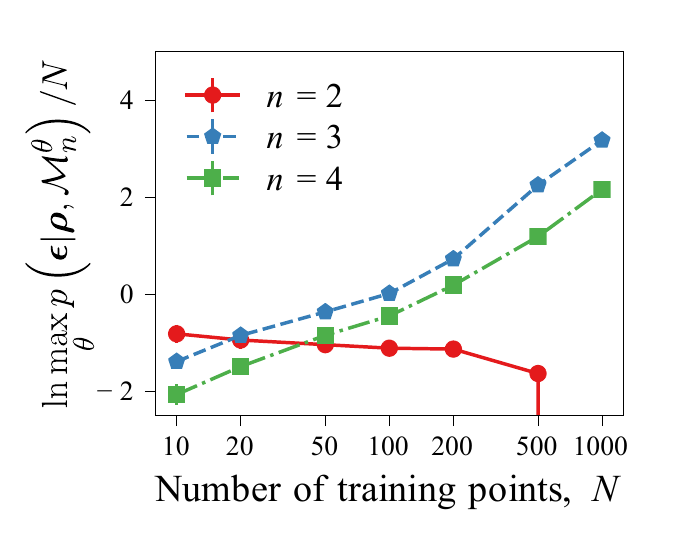}}
			\subfloat[$n^t = 4$\label{fig:ms_3c}]{\includegraphics[width=0.33\columnwidth]{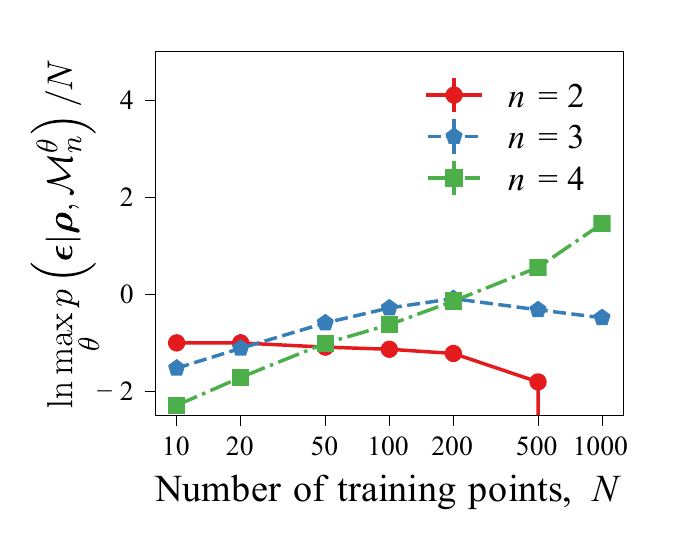}}
			\par\end{centering}
		\caption{Scaled log maximum marginal likelihood as a function of the number of training points for different kernel models $n$ and true interaction orders $n^t$.}
		\label{fig:scaled_log_marg_lik}
	\end{figure*}
	\par\end{center}

The results are reported in Figure \ref{fig:scaled_log_marg_lik}, where we graph the logarithm of the maximum marginal likelihood (MML), divided by the number of training points $N$, as a function of $N$ for different combinations of true orders $n^t$ and kernel order $n$.
The model selected in each case is the one corresponding to the line achieving the maximum value of this quantity.
It is interesting to notice that, when the kernels order is lower than the true order (i.e., for $n < n^t$), the MML can be observed to \emph{decreases} as a function of $N$ (as e.g., the red and blue lines in Figure \ref{fig:ms_3c}).
This makes the gap between the true model and the other models increase substantially as $N$ becomes sufficiently large.

Figure \ref{fig:model_selection_2} summarises the results of model selection. 
In particular, Figure \ref{fig:ms2a} illustrates the model-selected order $\hat{n}$ as a function of the true order $n^t$, for different training set sizes $N$. 
The graph reveals that, when the dataset is large enough ($N = 1000$ in this example) maximising the marginal likelihood always yields the true interaction order (green line). 
On the contrary, for smaller database sizes, a lower interaction order value $n$ is selected (blue and red lines).
This is consistent with the intuitive notion that smaller databases may simply not contain enough information to justify the selection of a complex model, so that a simpler one should be chosen.  
More insight can be obtained by observing Figure \ref{fig:ms2b}, reporting the model selected order as a function of the training dataset size for different true interaction orders. 
While the order of a simple 2-body model is always recovered (red line), to identify as optimal a higher order interaction model a minimum number of training points is needed, and this number grows with the system complexity.
Although not immediately obvious, choosing a simpler model when only limited databases are available also leads to smaller prediction errors on unseen configurations, since overfitting is ultimately prevented, as illustrated in Refs.~\cite{Glielmo:2017dj,Glielmo:2018bm} and further below in Section \ref{subsec:practice_1}.

\begin{center}
	\begin{figure*}
		\begin{centering}
			\subfloat[\label{fig:ms2a}]{\includegraphics[width=0.5\columnwidth]{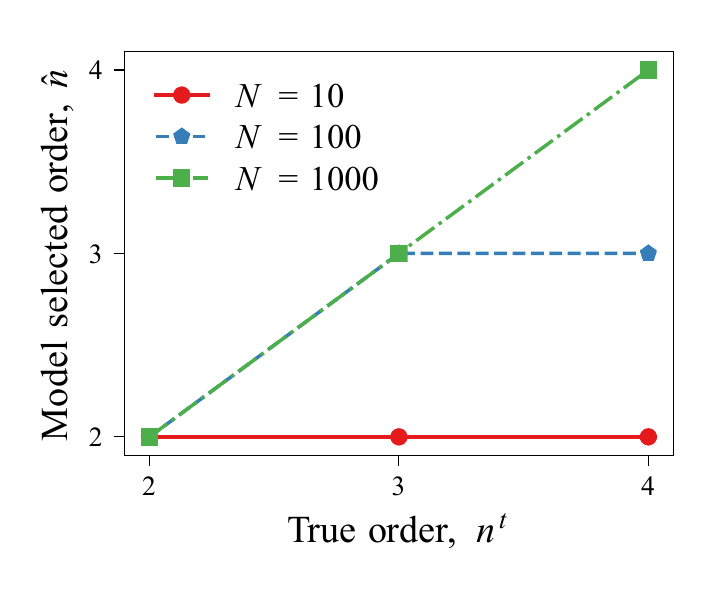}
				
			}\subfloat[\label{fig:ms2b}]{\includegraphics[width=0.5\columnwidth]{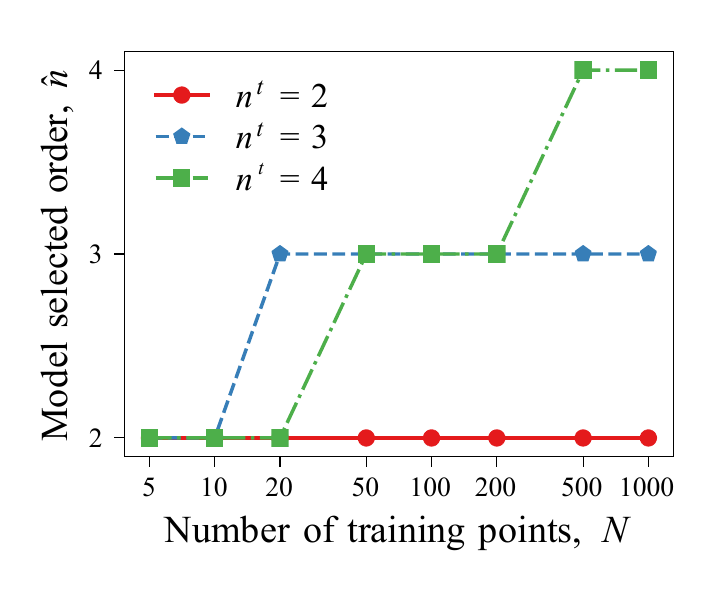}
			}
			\par\end{centering}
		
		\caption{Model selected order $\hat{n}$ as a function of the true order $n^t$ (left) and as a function of the number of training data points $N$ (right).}
		\label{fig:model_selection_2}
	\end{figure*}
	\par\end{center}

The picture emerging from these observations is one in which, although the quantum interactions occurring in atomistic systems will in principle involve all atoms in the system, there is never going to be sufficient data 
to select/justify the use of interaction models beyond the first few terms of the many-body expansion (or any similar expansion based on prior physical knowledge). 
At the same time, in many likely scenarios, a realistic target threshold for the average error on atomic forces (typically of the order of 0.1eV/A) will be met by truncating the series at a complexity order that is still practically manageable. 
Hence, in practice \emph{a small finite order model will always be optimal}. 

This is in stark contrast with the original hope of finding a single many-body \q{universal approximator} model to be used in every context, which has been driving a lot of interest in the early days of the ML-FF research field, producing for instance reference methods \cite{Behler:2007fe,Bartok:2010fj}. 
Furthermore, the observation that it may be possible to use models of finite-order complexity without ever recurring to universal approximators suggests alternative routes for increasing the accuracy of GP models without increasing the kernels' complexity. 
These are worth a small digression.

\begin{figure}[t]
	\sidecaption[t]
	\includegraphics[width=0.55\columnwidth]{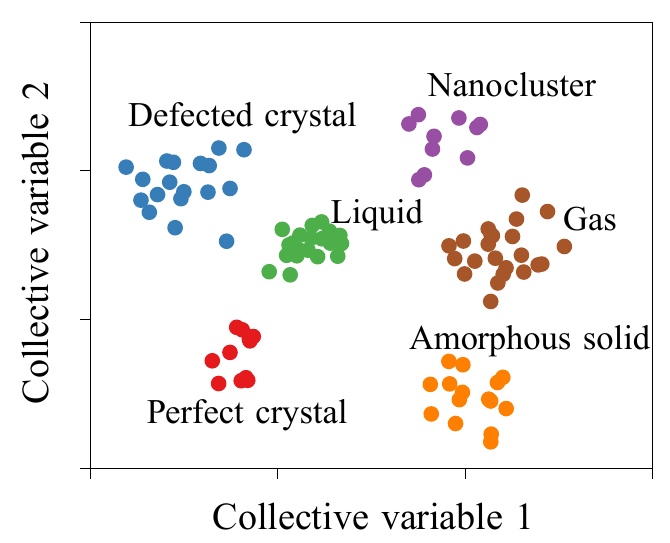}
	\caption{A an illustrative representation of an heterogeneous database composed of configurations which \q{cluster} around specific centroids in an arbitrary two dimensional space. The different clusters can be imagined to be different phases of the same material.}
	\label{fig:clusters}
\end{figure}

Imagine a situation as the one depicted in Figure \ref{fig:clusters}, where we have an heterogeneous dataset composed of configurations that cluster into groups.
This could be the case, for instance, if we imagine collecting a database which includes several relevant phases of a given material.
Given the large amount of data and the complexity of the physical interactions within (and between) several phases, we can imagine the model selected when training on the full dataset to be a relatively complex one.
On the other hand, each of the small datasets representative of a given phase may be well described by a model of much lower complexity.
As a consequence, one could choose to train several GP, one for each of the phases, as well as a \emph{gating function} $p(c \mid \rho)$ deciding, during an MD run, which of the clusters $c$ to call at any given time. 
These GPs learners will effectively \emph{specialise} on each particular phase of the material. 
This model can be considered a type of \emph{mixture of experts} model \cite{Jacobs:1991bk,Rasmussen:2002ty}, and heavily relies on a viable partitioning of the configuration space into clusters that will comprise similar entries.
This subdivision is far from trivially obtained in typical systems, and in fact obtaining \q{atlases} for real materials or molecules similar the one in Figure \ref{fig:clusters} is an active area of research \cite{De:2016ia, Ghiringhelli:2015kr, Mavracic:2018da,De:2017kx}. 
However, another simpler technique to combine multiple learner is that of bootstrap aggregating (\q{Bagging})~\cite{Breiman:1996fi}. 
In our particular case, this could involve training multiple GPs on random subsections of the data and then averaging them to obtain a final prediction.
While it should not be expected that the latter combination method will perform better than a GP trained on the full dataset, the approach can be very advantageous from a computational perspective since, similar to the mixture of experts model, it circumvents the $\mathscr{O}(N^3)$ computational bottleneck of inverting the kernel matrix in Eq.~\eqref{eq: energy_pred} by distributing the training data to multiple GP learners. 
ML algorithms based on the use of multiples learners belong to a broader class of \emph{ensemble learning} algorithms \cite{Sagi:2018cu,:2008eq}.

\subsection{Kernels for multiple chemical species}
\label{subsec:theory_multispecies_kernels}

In this section we briefly show how kernels for multispecies systems can be constructed, and provide specific expressions for the case of 2- and 3-body kernels. 

It is convenient to show the reasoning behind multispecies kernel construction starting from a simple example. 
Defining by $s_j$ the chemical species of atom $j$, a generic 2-body decomposition of the local energy of an atom $i$ surrounded by the configuration $\rho_i$ takes the form
\begin{equation}
\varepsilon(\rho_i) = \sum_{j \in \rho_i} \tilde{\varepsilon}_2^{s_is_j}(r_{ij}),
\end{equation}
where a pairwise function $\tilde{\varepsilon}_2^{s_is_j}(r_{ij})$ is assumed to provide the energy associated to each couple of atoms $i$ and $j$ which depends on their distance $r_{ij}$ and on their chemical species $s_i$ and $s_j$.
These pairwise energy functions should be invariant upon re-indexing of the atoms i.e., $\tilde{\varepsilon}_2^{s_is_j}(r_{ij}) = \tilde{\varepsilon}_2^{s_js_i}(r_{ji})$.
The kernel for the function $\varepsilon(\rho_i)$ then takes the form
\begin{equation}
\begin{split}
	k_2^s(\rho_i,\rho_l') & = \langle \varepsilon(\rho_i)\varepsilon(\rho_l') \rangle \\
	& = \sum_{jm} \langle \tilde{\varepsilon}_2^{s_is_j}(r_{ij}) \tilde{\varepsilon}_2^{s_l's_m'}(r_{lm}') \rangle \\
	& = \sum_{jm} \tilde{k}_2^{s_is_js_l's_m'}(r_{ij},r_{lm}'). \\
\end{split}
\end{equation}
The problem of designing the kernel $k_2^s$ for two configurations in this way reduced to that of choosing a suitable kernel $\tilde{k}_2^{s_i s_j s_l' s_m' }$ comparing couples of atoms.
An obvious choice for this would include a squared exponential for the radial dependence and a delta correlation for the dependence on the chemical species, giving rise to $\delta_{s_is_l'}\delta_{s_js_m'} k_{SE}(r_{ij},r_{lm}')$.
This kernel is however still not symmetric upon the exchange of two atoms and it would hence not impose the required property $\tilde{\varepsilon}_2^{s_is_j}(r_{ij}) = \tilde{\varepsilon}_2^{s_js_i}(r_{ji})$ on the learned pairwise potential. 
Permutation invariance can be enforced by a direct sum over the permutation group, in this case simply an exchange of the two atoms $l$ and $m$ in the second configuration.
The resulting 2-body multispecies kernel reads

\begin{equation}
	k_2^s(\rho_i,\rho_l') = \sum_{\substack{j \in \rho \\ m \in \rho'}} (\delta_{s_i s_l'} \delta_{s_js_m'} + \delta_{s_i s_m'} \delta_{s_js_l'})\mathrm{e}^{-(r_{ij}-r_{lm}')^2/2\ell^2}.
\end{equation}
This can be considered the natural generalisation of the single species 2-body kernel in Eq.~(\ref{eq:2bodyk}).
A very similar sequence of steps can be followed for the 3-body kernel. 
By defining the vector containing the chemical species of an ordered triplet as \mbox{$\mathbf{s}_{ijk} = (s_i s_j s_k)^{\rm{T}}$}, as well as the vector containing the corresponding three distances $\mathbf{r}_{ijk} = (r_{ij} r_{jk} r_{ki} )^{\rm{T}}$, a multispecies 3-body kernel can be compactly written down as
\begin{equation}
k_{3}^s(\rho_{i},\rho_{l}') 
= \sum_{\substack{j > k\in\rho_i \\ m > n\in\rho'_l}} \, 
\sum_{\mathbf{P} \in \mathscr{P}} \delta_{\mathbf{s}_{ijk},\mathbf{P}\mathbf{s}_{lmn}'}
\mathrm{e}^{-\| \mathbf{r}_{ijk}^{\rm{T}}-\mathbf{P} \mathbf{r}_{lmn}' \|^2/2 \ell^2},
\end{equation}
where the group $\mathscr{P}$ contains six permutations of three elements, represented by the matrices $\mathbf{P}$.
The above can be considered the direct generalisation of the 3-body kernel in Eq.~(\ref{eq:3bodyk}). 
It is simple to see how the reasoning can be extended to an arbitrary $n$-body kernel. 
Importantly, the computational cost of evaluating the multispecies kernels described above \emph{does not} increase with the number of species present in a given environment, and the kernels' interaction order could be increased arbitrarily at no extra computational cost using Eqs.~(\ref{eq:non_unique_kernel}) and (\ref{eq:k_nu_SE}).

\subsection{Summary}
In this section, we first went trough the basics of GP regression, and emphasised the importance of a careful design of the kernel function, which ideally should encode any available prior information on the (energy or force) function to be learned (Section \ref{subsec:theory_1}).
In Section \ref{subsec:theory_local_energies_from_forces} we detailed how a local energy function (which is not a quantum observable) can be learned in practice starting from a database containing solely total energies and atomic forces.
We then discussed how fundamental properties of the target force field, such as 
the interaction order, smoothness, as well as its permutation, translation and rotation symmetries, can be included into the kernel function (Section \ref{subsec:theory_prior}).
We next proceeded to the construction of a set of computationally affordable kernels that implicitly define smooth, fully symmetric potential energy functions with tunable \q{complexity} given a target interaction order $n$.
In Section \ref{subsec:theory_model_selection} we looked at the problem of choosing the order $n$ best suited for predictions based on the information available in a given set of QM calculations. 
Bayesian theory for model selection prescribes in this case to choose the $n$-kernel yielding the largest marginal likelihood for the dataset, which is found to work very well in a 1D model system where the interaction order can be tuned and is correctly identified upon sufficient training.  
Finally, in Section \ref{subsec:theory_multispecies_kernels} we showed how the ideas presented can be generalised to systems containing more than one chemical species.

\section{Practical considerations}
\label{sec:practice}

We next focus on the application of the techniques described in the previous sections.
In Section \ref{subsec:practice_1} we apply the model selection methodology described in Section \ref{subsec:theory_model_selection} to two atomic systems described using density functional theory (DFT) calculations. 
Namely, we consider a small set of models with different interaction order $n$, and recast the optimal model selection problem into an optimal kernel order selection problem.
This highlights the connections between the optimal kernel order $n$ and the physical properties of the two systems, revealing how novel physical insight can be gained via model selection.
We then present a more heuristic approach to kernel order selection and compare the results with the ones obtained from the MML procedure. 
The comparison reveals that typically the kernel selected via the Bayesian approach also incurs into lower average error for force prediction on a provided test set.
In Section \ref{subsec:practice_2} we discuss computational efficiency of GPs.
We argue that an important advantage of using GP kernels of known finite order is the possibility of \q{mapping} the kernel's predictions onto the values of a compact approximator function of the same set of variables. 
This keeps all the advantages of the Bayesian framework, while removing the need of lengthy sums over the database and expensive kernel evaluations typical of GP predictions.
For this we introduce a method that can be used to \q{map} the GP predictions for finite-body kernels and therefore increase the computational speed up to a factor of 10$^4$ when compared with the original $3$-body kernel, while effectively producing identical interatomic forces.  

\subsection{Applying model selection to nickel systems} %
\label{subsec:practice_1}	

\begin{center}
	\begin{figure*}
		\begin{centering}
			\subfloat[\label{fig:nickel_systems}]{\includegraphics[width=0.5\columnwidth]{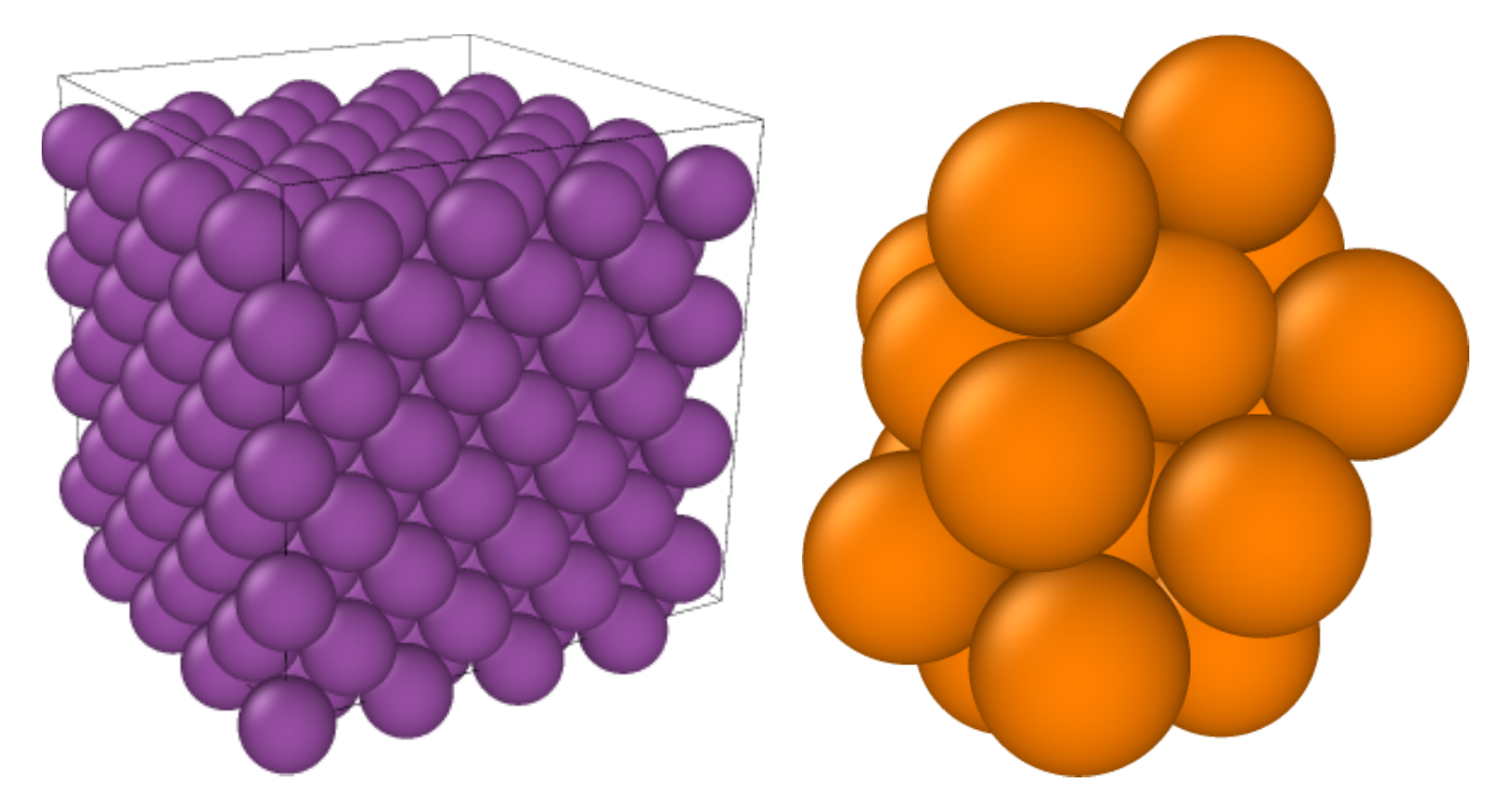}
				
			}\subfloat[\label{fig:ML_on_nickel}]{\includegraphics[width=0.5\columnwidth]{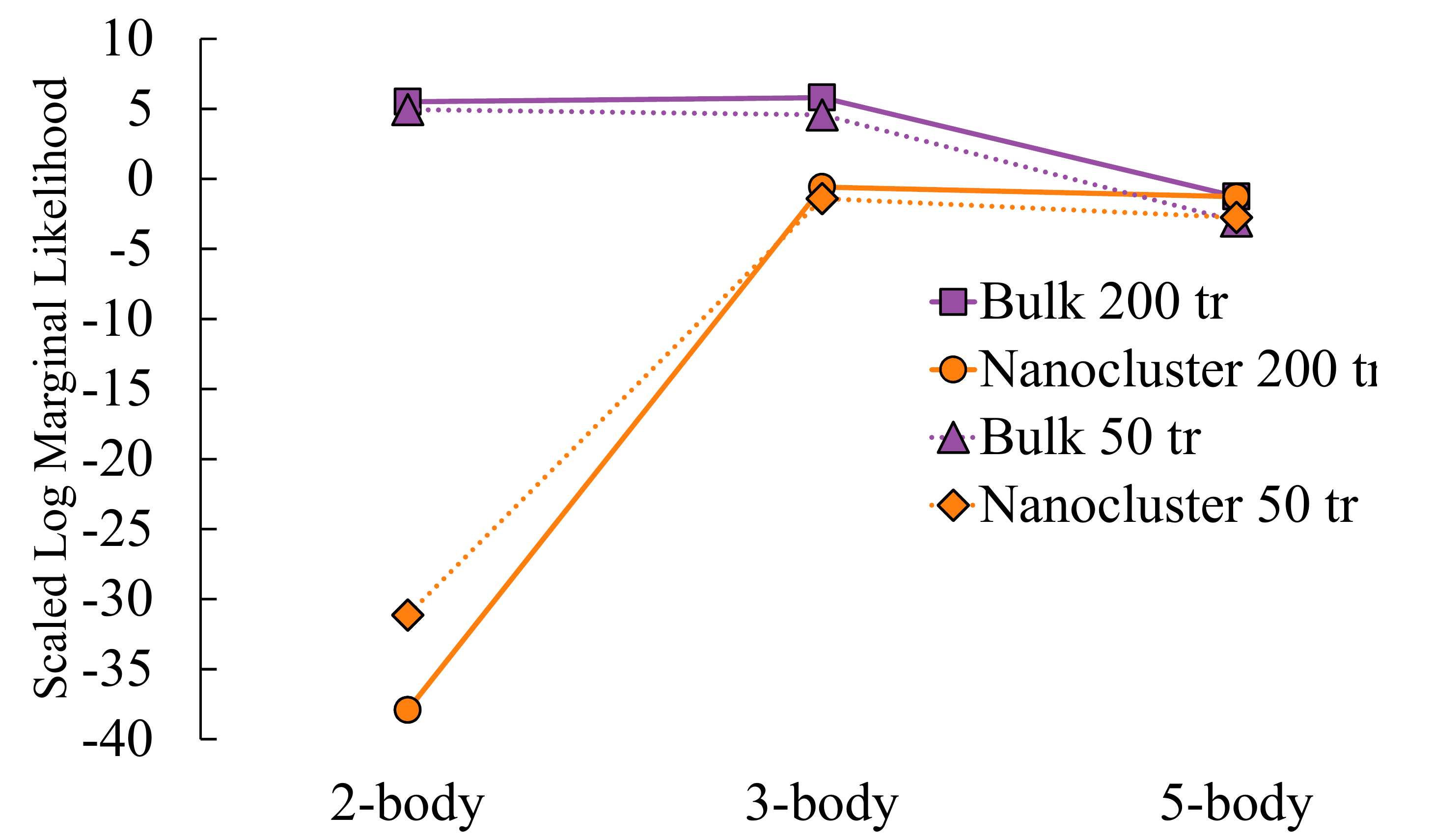}
			}
			\par\end{centering}
		
	\caption{Panel (a): the two Nickel systems used in this section as examples, with bulk FCC Nickel in periodic boundary conditions on the left (purple) and a Nickel nanocluster containing 19 atoms on the right (orange). Panel (b): maximum log marginal likelihood divided by the number of training points for the 2-, 3- and 5-body kernels in the bulk Ni (purple) and Ni nanocluster (orange) systems, using 50 (dotted lines) and 200 (full lines) training configurations.} 
	\end{figure*}
	\par\end{center}

We consider two Nickel systems: a bulk face centered cubic (FCC) system described using periodic boundary conditions (PBC), and a defected double icosahedron nanocluster containing 19 atoms, both depicted in Figure \ref{fig:nickel_systems}.
We note that all atoms in the bulk system experience a similar environment, their local coordination involving 12 nearest neighbours,  
as the system contains no surfaces, edges or vertexes. 
The atom-centred configurations $\boldsymbol{\rho}$ are therefore very similar in this system. 
The nanocluster system is instead exclusively composed by surface atoms, involving a different number of nearest neighbours for different atoms. 
The GP model is thus here required to learn the reference force field for a significantly more complex and more varied set of of configurations. 
It is therefore expected that the GP model selected for the nanocluster systems will be more complex (have a higher kernel order $n$) than the one selected for the bulk system, even if the latter system is kept at an appreciably higher temperature.

The QM databases used here were extracted from first principles MD simulations carried out at $500\rm{K}$ in the case of bulk Ni, and at $300\rm{K}$ for the Ni nanocluster.
All atoms within a $4.45\text{\AA}$ cutoff from the central one were included in the atomic configurations $\boldsymbol{\rho}$ for the bulk Ni system, while no cutoff distance was set for the nanocluster configurations, which therefore all include 19 atoms. 
In this example we perform model selection on a a restricted, yet representative, model set $\{\mathscr{M}_2^{\boldsymbol{\theta}}, \mathscr{M}_3^{\boldsymbol{\theta}}, \mathscr{M}_5^{\boldsymbol{\theta}} \}$ containing, in increasing order of complexity, a 2-body kernel (see Eq. (\ref{eq:2bodyk})), a 3-body kernel (see Eq. (\ref{eq:3bodyk})), and a non unique 5-body kernel obtained by squaring the 3-body kernel 
\cite{Glielmo:2017dj} (see Eq. (\ref{eq:non_unique_kernel})).
Every kernel function depends on only two hyperparameters $\boldsymbol{\theta} = (\ell, \sigma_n)$, representing the characteristic lengthscale of the kernel $\ell$ and the modelled uncertainty of the reference data $\sigma_n$ .
While the value of $\sigma_n$ is kept the same for all kernels, we optimise the lengthscale parameter $\ell$ for each kernel via marginal likelihood  maximisation (Eq.~(\ref{eq: ML})).
We then select the optimal kernel order $n$ as the one associated with the highest marginal likelihood.

Figure \ref{fig:ML_on_nickel} reports the optimised marginal likelihood of the three models ($n=2,3,5$) for the two systems while using 50 and 200 training configurations.
The 2- and 3-body kernels reach comparable marginal likelihoods in the bulk Ni system, while a 3-body kernel is instead always optimal for the Ni nanocluster system. 
While intuitively correlated with the relative complexity of the two systems, these results yield further interesting insight.
For instance, the occurrence of angular-dependant forces must have a primary role in small Ni clusters since a 3-body kernel is necessary and sufficient to accurately describe the atomic forces in the nanocluster. 
Meanwhile, the 5-body kernel does not yield a higher likelihood, suggesting that the extra correlation it encodes is not significant enough to be resolved at this level of training. 
On the other hand, the forces on atoms occurring in a bulk Ni environment at a temperature as high as 500K are well described by a function of radial distance only, suggesting that angular terms play little to no role, as long as the bonding topology remains everywhere that of undefected FCC crystal. 

The comparable maximum log marginal likelihoods the 2- and 3-body kernels produce on bulk environment suggest that the two kernels will achieve similar accuracies.
In particular the 2-body kernel produces the higher log marginal likelihood when the models are trained using $N$ = 50 configurations, while the 3-body kernel has a better performance when $N$ increases to 200.
This result resonates with the results shown on the toy model in section \ref{fig:model_selection_2}: the model selected following the MML principle is a function of the number of training points $N$ used.
For this reason, when using a restricted data set we should prefer the 2-body kernel to model bulk Ni and a 3-body kernel to model the Ni cluster, as these provide the simplest models that are able to capture sufficiently well the interactions of the two systems.
Notice that the models selected in the two cases are different and this reflects the different nature of the chemical interactions involved.
This is reassuring, as it shows that the MML principle is able to correctly identify the minimum interaction order needed for a fundamental characterisation of a material even with very moderate training set sizes.
For most inorganic material this minimum order can be expected to be low (typically either 2 or 3) as a consequence of the ionic or covalent nature of the chemical bonds involved, while for certain organic molecules, one can expect this to be higher (think e.g., at the important of 4-body dihedral terms).

Overall, this example showcases how the maximum marginal likelihood principle can be used to automatically select the simplest model which accurately describes the system, meanwhile providing some insight on the nature of the interactions occurring in the system.
In the following, we will compare this procedure with a more heuristic approach based on comparing the kernels' generalisation error, which is  commonly employed in the literature \cite{ Glielmo:2017dj, Glielmo:2018bm, Zeni:2018to, huo2017unified, kruglov2017energy} for its ease of use.

\begin{center}
	\begin{figure*}
		\begin{centering}
			\subfloat[\label{fig:learning_curves_a}]{\includegraphics[width=0.5\columnwidth]{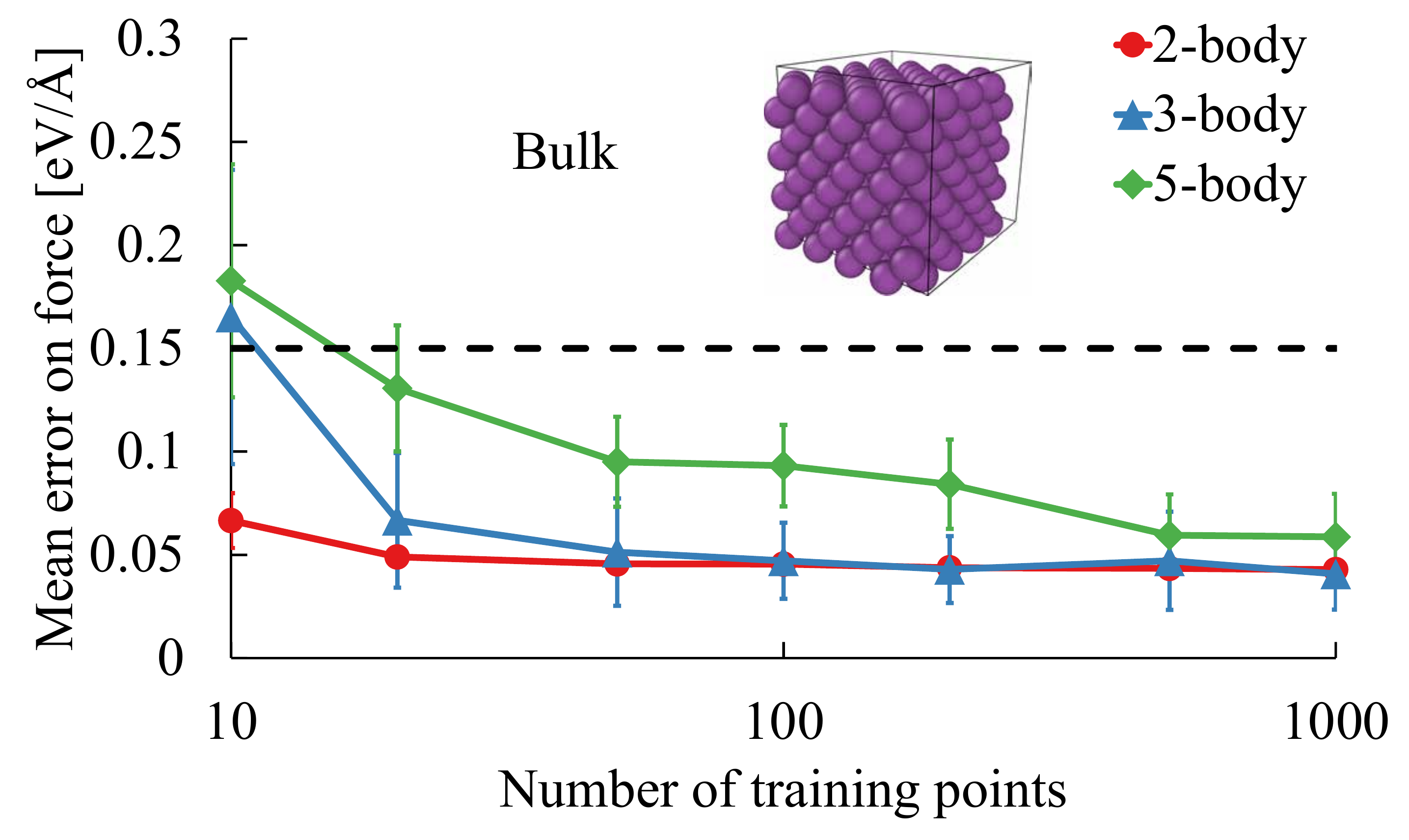}
				
			}\subfloat[\label{fig:learning_curves_b}]{\includegraphics[width=0.5\columnwidth]{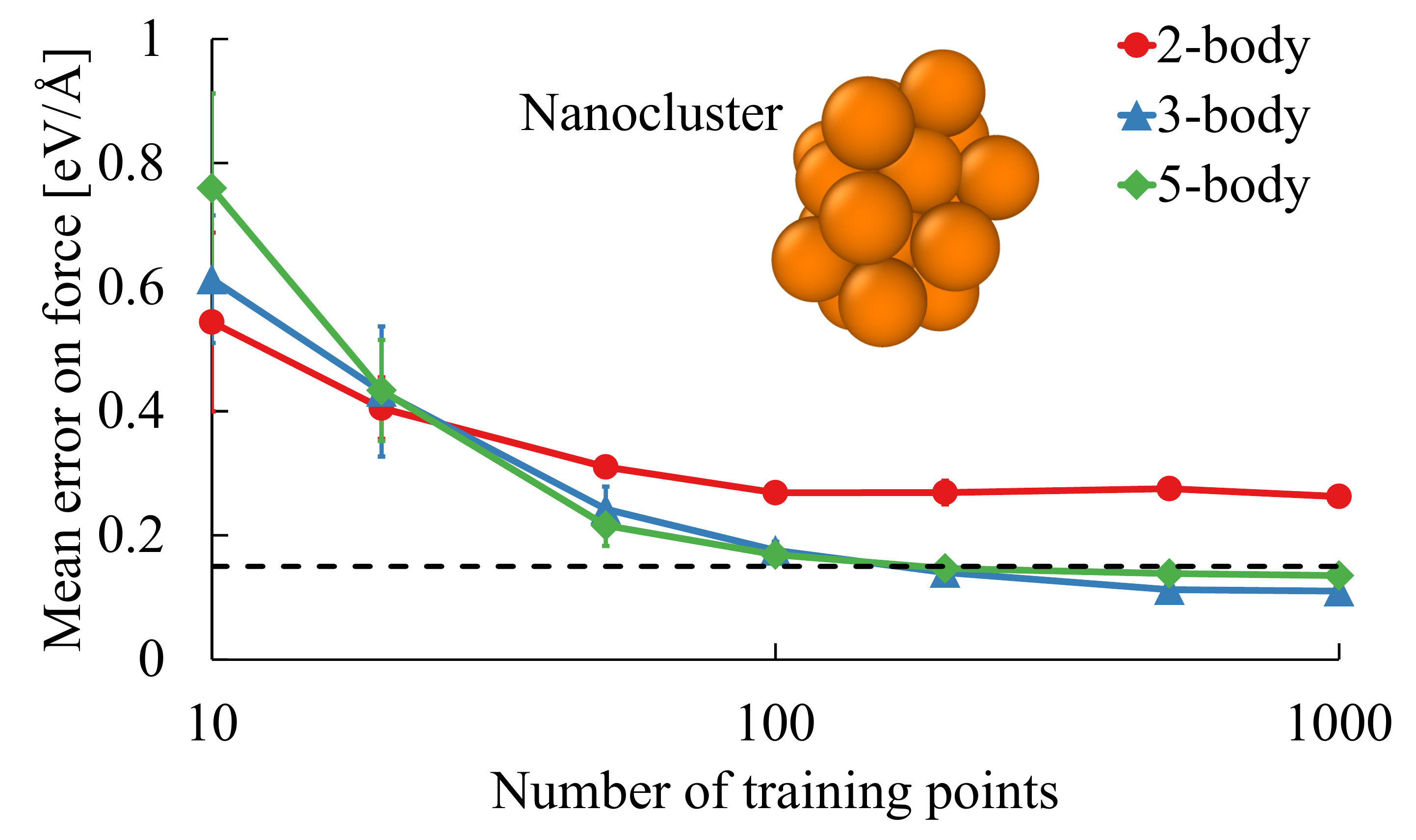}
			}
			\par\end{centering}
		
	\caption{Learning curves for Bulk Ni (a) and Ni nanocluster (b) systems displaying the mean error incurred by the 2-body, 3-body and 5-body kernels as the number of training points used varies. The \q{error on force} reported here is defined as the norm of the difference vector between predicted and reference force. The error bars in the graphs show the standard deviation when 5 tests were repeated using different randomly chosen training and testing configuration sets. The black dashed line correspond to the same target accuracy in the two cases (here 0.15 eV/$\text{\AA}$), much more easily achieved in the bulk system.}
	\label{fig:learning_curves} 
	\end{figure*}
	\par\end{center}

Namely, let us assume that all of the hyperparameters $\boldsymbol{\theta}$ have been optimized for each kernel in our system of interest, either via maximum likelihood optimization or via manual tuning.
We then measure the error incurred by each kernel on a \emph{test set} i.e., a set of randomly chosen configurations and forces different from those used to train the GP.
Tracing this error as the number of training points increases, we obtain a learning curve (Figure \ref{fig:learning_curves}).
The selected model will be the lowest-complexity one that is capable of reaching a target accuracy (chosen by the user, here set to 0.15 eV/$\text{\AA}$, cf. black dotted line in Figure \ref{fig:learning_curves}). 
Since lower-complexity kernels are invariably faster learners, if they can reach the target accuracy they will do so using a smaller number of training points, consistent with all previous discussions and findings. 
More importantly, lower complexity kernels are computationally faster and more robust extrapolators than higher-complexity ones - a property that derives from the low order interaction they encode.
Furthermore, they can be straightforwardly mapped as described in the next section.
For the bulk Ni system of the present example, all three kernels reach the target error threshold, so the 2-body kernel is the best choice for the bulk Ni system.
In the Ni nanocluster case the 2-body kernel is not able to capture the complexity of force field experienced by the atoms in the system, while both the 3- and 5-body kernels reach the threshold. 
Here the 3-body kernel is thus preferred.

In conclusion, marginal likelihood and generalisation error offer different approaches to the problem of optimal model selection. 
While their outcomes are generally consistent, they two methods differ in spirit e.g. because  the marginal likelihood distribution naturally incorporates information on the underlying model's variance when measured on the training target data and this will reflect into selecting the best model also on this basis (see Fig. \ref{fig:model_selection}, in which the target data $\epsilon_0^r$ select the model with $n=3$). 
This is not true when using the generalisation error, where all that counts is the model's prediction, i.e., the predicted mean of the posterior GP. 
Moreover, while model selection according to the marginal likelihood is a function of the training set only, the generalisation error is also dependant on the choice of the test set, whose sampling uncertainty can be reduced through repeated tests, as reported in Fig. \ref{fig:learning_curves}.
Regardless of the model selection method, simpler models may perform better when the available data is limited, i.e., higher model complexity does not necessarily imply higher prediction accuracy:
whether this is the case will each time depend on the target physical system, the desired accuracy threshold, and the amount data available for training.
Due to the lower dimensionality of the feature spaces used to construct the kernels, the predictions of simpler models will also be easier to re-express into a more computationally efficient way than carrying out the summation in Eq. (\ref{eq:explicit_GP_pred}). 
For the examples described in this chapter, this means re-expressing the trained GPs based on $n$-body kernels as functions of 3$n$-6 variables which can be evaluated directly, without using a database. 
These functions can be viewed as the nonparametric $n$-body classical force fields (here named \q{MFFs}) that the $n$-body kernels' predictions {\em exactly} correspond to. 
Exploiting this correspondence allows us to achieve force fields as fast-executing as determined by the complexity of the physical problem at hand (which will determine the lowest $n$ that can be used).
Examples of MFF constructions and tests on their computational efficiency are provided in the next section.

\subsection{Speeding up predictions by building MFFs}
\label{subsec:practice_2}

In Section \ref{subsec:theory_building_nkernels} we described how simple $n$-body kernels of any order $n$ could be constructed. 
Force prediction based on these kernels effectively produces non-parametric classical $n$-body force fields: typically depending on  distances (2-body) as well as on angles (3-body), dihedrals (4-body) and so on, but not bound by design to any particular functional form.

In this section we describe a mapping technique (first presented in Ref. \cite{Glielmo:2018bm}) that faithfully encodes  forces produced by $n$-body GP regression into classical tabulated force fields. 
This procedure can be carried out with arbitrarily low accuracy loss, and always yields a substantial computational speed gain. 

\begin{center}
	\begin{figure*}
		\begin{centering}
			\subfloat[\label{fig:mapping_error}]{\includegraphics[width=0.5\columnwidth]{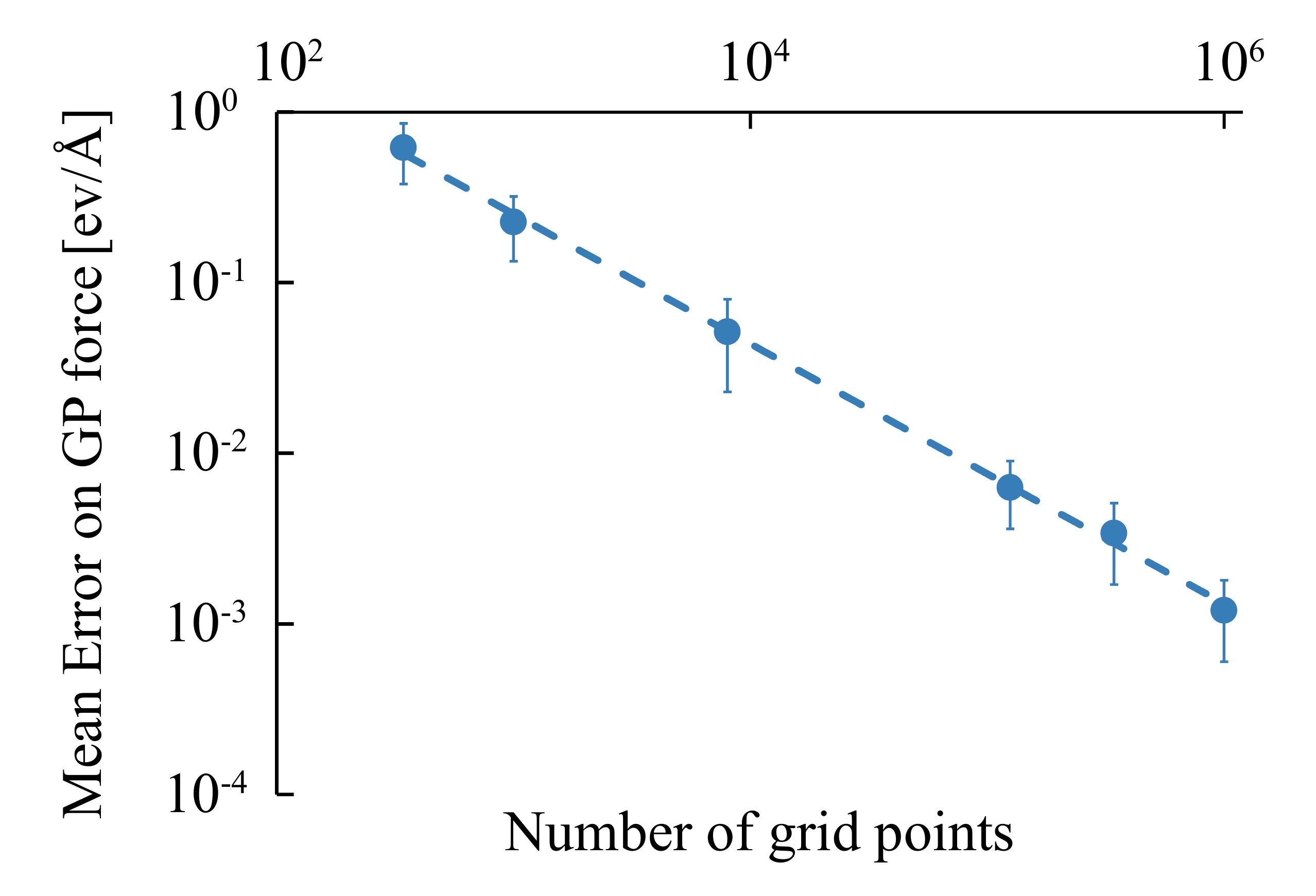}
				
			}\subfloat[\label{fig:mapping_speed}]{\includegraphics[width=0.5\columnwidth]{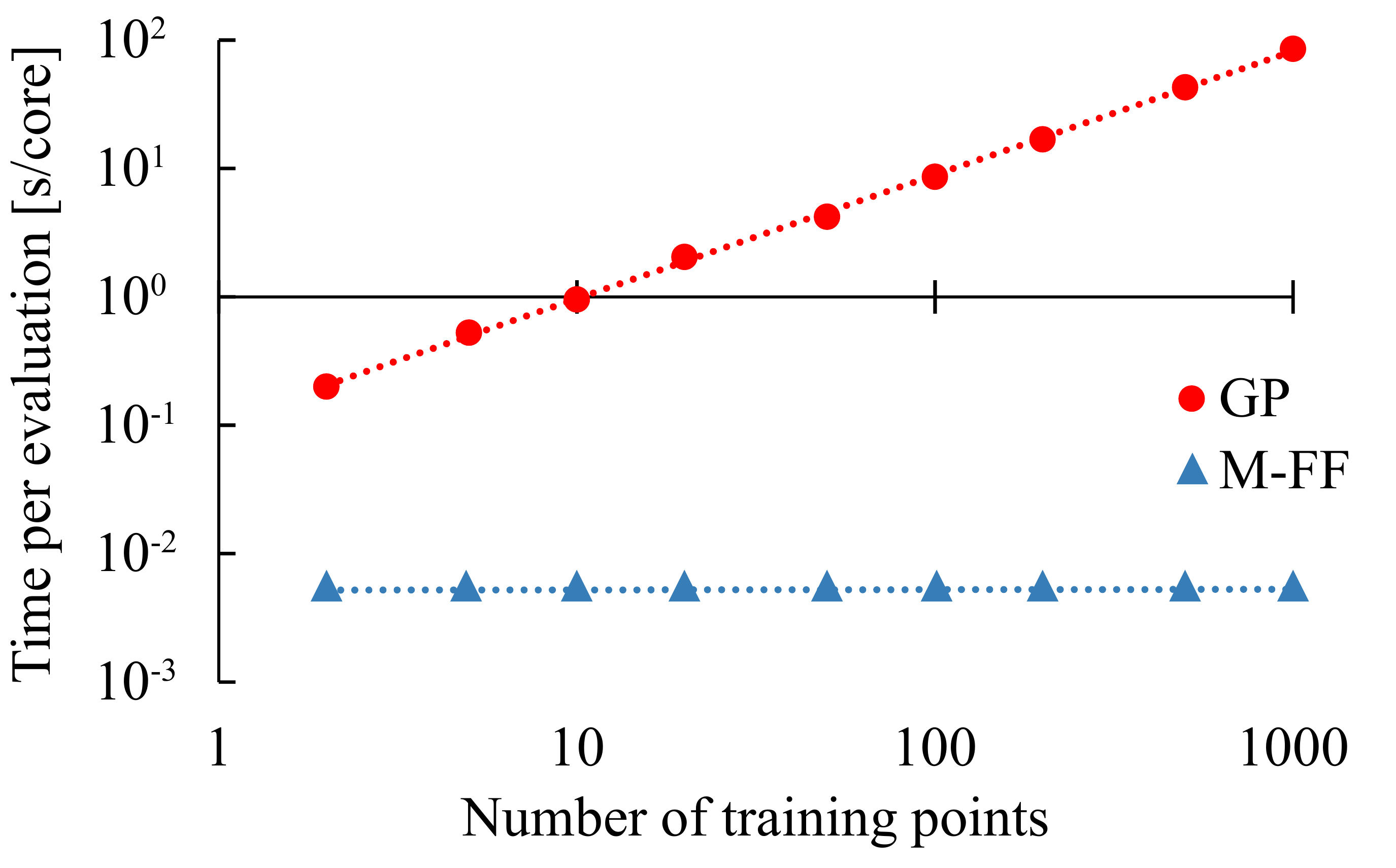}
			}
			\par\end{centering}		
\caption{Panel (a): error incurred by a 3-body MFF w.r.t. the predictions of the original GP used to build it as a function of the number of points in the MFF grid.
Panel (b): computational time needed for the force prediction on an atom in a 19-atoms Ni nanocluster as a function of the number of training points for a 3-body GP (red dots) and for the MFF built from the same 3-body GP (blue dots).
	} 
\end{figure*}
\par\end{center}

We start from the expression of the GP energy prediction in Eq.~(\ref{eq:explicit_GP_pred}), where we substitute $k$ with a specific $n$-body kernel (in this example, the 2-body kernel of Eq. (\ref{eq: k2}) for simplicity). 
Rearranging the sums we obtain:
\begin{equation}
\hat{\epsilon}(\rho) = \sum_{i \in \rho} \left( \sum_d^{N} \sum_{j \in \rho_d} e^{-(r_{i}-r_{j})^2/2\ell^2} \alpha_d \right).
\label{eq: 2body_rearranged}
\end{equation}
The expression within the parentheses in the above equation is a function of the single distance $r_i$ in the target configuration $\rho$ and the training dataset, and it will not change once the dataset is chosen and the model is trained (the covariance matrix is computed and inverted to give the coefficient $\alpha_d$ for each dataset entry).  
We can thus rewrite Eq. (\ref{eq: 2body_rearranged}) as
\begin{equation}
\hat{\epsilon}(\rho) = \sum_{i \in \rho} \tilde{\epsilon}_2(r_i),
\label{eq: mapped_2body}
\end{equation}
where the function $\tilde{\epsilon}(r_i)$ can be now thought to be nonparametric 2-body potential expressing the energy associated to an atomic pair (a \q{bond}) as a function of the interatomic distance, so that the energy associated with a local configuration $\rho$ is simply the sum over all atoms surrounding the central one of this 2-body potential.
It is now possible to compute the values of $\tilde{\epsilon}_2(r_i)$ for a set of distances $r_i$, store them in an array, and from here on \textit{interpolate} the value of the function for any other distances rather than using the GP to compute this function for every atomic configuration during an MD simulation.
In practice, a spline interpolation of the so-tabulated potential can be very easily used to predict any $\hat{\epsilon}(\rho)$ or its negative gradient  $\hat{\mathbf{f}}(\rho)$ (analytically computed, to allow for a constant of motion in MD runs).
The interpolation approximates the GP predictions with arbitrary accuracy, which increases with the density of the grid of tabulated values, as illustrated in Figure \ref{fig:mapping_error}. 

The computational speed of the resulting \q{mapped force field} (MFF) is independent of the number of training points $N$ and depends linearly, rather than quadratically, on the number of distinct atomic $n$-plets present in a typical atomic environment $\rho$ including $M$ atoms plus the central one (this is the number of combinations $\binom{M}{n-1}= M!/(n-1)!(M-n+1)!$, yielding e.g., $M$ pairs and $M(M-1)/2$ triplets).
The resulting overall $N \binom{M}{n-1} $ speedup factor is typically several orders of magnitude over the original $n$-body GP, as shown in Figure \ref{fig:mapping_speed}.\\

The method just described can in principle be used to obtain $n$-body MFFs from any $n$-body GPs, for every finite $n$. 
In practice however, while mapping 2-body or 3-body predictions on a 1D or 3D spline is straightforward, the number of values to store grows exponentially for $n$, consistent with the rapidly growing dimensionality associated with atomic $n$-plets. 
This makes the procedure quickly not viable for higher $n$ values which would require (3$n$-6)-dimensional mapping grids and interpolation splines. 
On a brighter note, flexible 3-body force fields were shown to capture most of the features for a variety of inorganic materials
\cite{Glielmo:2018bm, Zeni:2018to, Takahashi:2017th, Deringer:2017ea}. 
Increasing the order of the kernel function beyond 3 might be unnecessary for many systems (and if only few training data are available, it could be still advantageous to use a low--$n$ model to \textit{improve} prediction accuracy, as discussed in Section \ref{subsec:theory_model_selection}).
MFFs cab be built for systems containing any number of atomic species. 
As already described in Section \ref{subsec:theory_multispecies_kernels}, the cost of constructing a multispecies GP does not increase with the number of species modelled. 
On the other hand, the number of $n$-body MFFs that need to be constructed when $k$ atomic species are present grows as the multinomial factor $\frac{(k + n - 1)!}{n!(k-1)!}$ (just as any classical force field of the same order).
Luckily, constructing multiple MFFs is an embarrassingly parallel problem as different MFFs can be assigned to different processors.
This means that the MFF construction process can be considered affordable also for high values of $k$, especially when using a 3-body model (which can be expected to achieve sufficient accuracy for a large number of practical applications).

We finally note that the variance of a prediction $\hat{\sigma}^2 (\rho)$ (third term in Eq.~(\ref{eq: energy_pred})) could also be mapped similarly to its mean.
However, it is easy to check that the mapped variance will have twice as many arguments as the mapped mean, which again makes the procedure rather cumbersome for $n>2$.
For instance, for $n=2$ one would have to store the function of two variables $\tilde{\sigma}^2(r_i,r_j)$ providing the variance contribution from any two distances within a configuration, and the final variance can be computed as a sum over all contributions.
A more affordable estimate of the error could also be obtained by summing up only the contributions coming from single $n$-plets (i.e., $\tilde{\sigma}^2(r_i,r_i)$ in the $n=2$ example). This alternative measure could again be mapped straightforwardly also for $n=3$ and its accuracy in modelling the uncertainty in the real materials should be investigated. 

\vspace{0.2cm}

MFFs obtained as described above have already been used to perform MD simulations on very long timescales while tracking with very good accuracy their reference \emph{ab-initio} DFT calculations for a set of Ni$_{19}$ nanoclusters \cite{Zeni:2018to}.
In this example application, a total of  $1.2\cdot10^8$ MD time steps were typically required 24 CPUs in $\sim 3.75$ days. 
The same simulation would have taken $\sim 80$ years before mapping, and indicatively 
$\sim 2000$ years using the full DFT-PBE (Perdew-Burke-Ernzerhof) spin-orbit coupling method which was used to build the training database.
A Python implementation for training and mapping two and three body nonparametric force fields for one or two chemical species is freely available within the MFF package \cite{mff_package}.

\section{Conclusions}
\label{sec:4}
In this chapter we introduced the formalism of Gaussian process regression for the construction of force fields.
We analysed a number of relevant properties of the kernel function, namely its smoothness and its invariance with respect to permutation of identical atoms, translation, and rotation.
The concept of interaction order, traditionally useful in constructing classical parametrised force fields and recently imported into the context of machine learning force fields, was also discussed.
Examples on how to construct smooth and invariant $n$-body energy kernels have been given, with explicit formulas for the cases of $n=2$ and $n=3$.
We then focused on the Bayesian model selection approach, which prescribes the maximisation of the marginal likelihood, and applied it to a set of standard kernels defined by and integer order $n$.
In a 1D system where the target interaction order could be exactly set, explicit calculations exemplified how the optimal kernel order choice depends on the number of training points used, so that larger datasets are typically needed to resolve the appropriateness of more complex models to a target physical system.\\
We next reported an example of application of the marginal likelihood maximisation approach to kernel order selection for two Nickel systems: face centred cubic crystal and a Ni$_{19}$ nanocluster.
In this example, prior knowledge about the system provides hints on the optimal kernel order choice which is \emph{a posteriori} confirmed by the model selection algorithm based on the maximum marginal likelihood strategy.
To complement the Bayesian approach to kernel order selection, we briefly discussed the use of learning curves based on the generalisation error to select the simplest model that reaches a target accuracy.
We finally introduced the concept of \q{mapping} GPs onto classical MFFs, and exemplified how mapping of mean and variance of a GP energy prediction can be carried out, providing explicit expressions for the case of a 2-body kernel.
The construction of MFFs allows for an accurate calculation of GP predictions while reducing the computational cost by a factor $\sim 10^4$ in most operational scenarios of interest in materials science applications, allowing for molecular dynamics simulations that are as fast as classical ones but with an accuracy that approaches \emph{ab-initio} calculations.
\begin{acknowledgement}

The authors acknowledge funding by the Engineering
and Physical Sciences Research Council (EPSRC) through the Centre for Doctoral Training “Cross Disciplinary
Approaches to Non-Equilibrium Systems” (CANES, Grant No. EP/L015854/1) and by the Office of Naval Research Global (ONRG Award No. N62909-15-1-N079). 
The authors thank the UK Materials and Molecular Modelling Hub for computational resources, which is partially funded by EPSRC (EP/P020194/1).
ADV acknowledges further support by the
EPSRC HEmS Grant No. EP/L014742/1 and by the
European Union’s Horizon 2020 research and innovation
program (Grant No. 676580, The NOMAD Laboratory,
a European Centre of Excellence). 
\vspace{0.1cm}

  We, AG, CZ, and AF, are immensely grateful to Alessandro De Vita for having devoted, with inexhaustible energy and passion, an extra-ordinary amount of his time and brilliance towards our personal and professional growth.
 
\end{acknowledgement}

\bibliographystyle{spphys}
\bibliography{newFULL,FULL_2}

\end{document}